\documentstyle{mn}
\input{psfig}

\newif\ifAMStwofonts

\newcommand{\tP}{\tilde{P}}
\newcommand{\tF}{\tilde{F}}

\newcommand{\tA}{\tilde{A}}
\newcommand{\tB}{\tilde{B}}

\newcommand{\tx}{\tilde{x}}

\renewcommand{\bar}{\overline }

\newcommand{\xiav}{\bar{\xi}}

\ifoldfss
  \ifCUPmtlplainloaded \else
    \NewTextAlphabet{textbfit} {cmbxti10} {}
    \NewTextAlphabet{textbfss} {cmssbx10} {}
    \NewMathAlphabet{mathbfit} {cmbxti10} {} 
    \NewMathAlphabet{mathbfss} {cmssbx10} {} 
  \fi
  \ifAMStwofonts
    \ifCUPmtlplainloaded \else
      \NewSymbolFont{upmath} {eurm10}
      \NewSymbolFont{AMSa} {msam10}
      \NewMathSymbol{\upi}     {0}{upmath}{19}
      \NewMathSymbol{\umu}     {0}{upmath}{16}
      \NewMathSymbol{\upartial}{0}{upmath}{40}
      \NewMathSymbol{\leqslant}{3}{AMSa}{36}
      \NewMathSymbol{\geqslant}{3}{AMSa}{3E}

      \let\leq=\leqslant \let\le=\leqslant
      \let\geq=\geqslant \let\ge=\geqslant
    \fi
  \fi
\fi 

\ifnfssone
  \newmathalphabet{\mathit}
  \addtoversion{normal}{\mathit}{cmr}{m}{it}
  \addtoversion{bold}{\mathit}{cmr}{bx}{it}
  \newmathalphabet{\mathbfit} 
  \addtoversion{normal}{\mathbfit}{cmr}{bx}{it}
  \addtoversion{bold}{\mathbfit}{cmr}{bx}{it}
  \newmathalphabet{\mathbfss} 
  \addtoversion{normal}{\mathbfss}{cmss}{bx}{n}
  \addtoversion{bold}{\mathbfss}{cmss}{bx}{n}
  \ifAMStwofonts
    \ifCUPmtlplainloaded \else
      \UseAMStwoboldmath
      \makeatletter
      \new@mathgroup\upmath@group
      \define@mathgroup\mv@normal\upmath@group{eur}{m}{n}
      \define@mathgroup\mv@bold\upmath@group{eur}{b}{n}
      \edef\UPM{\hexnumber\upmath@group}
      \new@mathgroup\amsa@group
      \define@mathgroup\mv@normal\amsa@group{msa}{m}{n}
      \define@mathgroup\mv@bold\amsa@group{msa}{m}{n}
      \edef\AMSa{\hexnumber\amsa@group}
      \makeatother
      \mathchardef\upi="0\UPM19
      \mathchardef\umu="0\UPM16
      \mathchardef\upartial="0\UPM40
      \mathchardef\leqslant="3\AMSa36
      \mathchardef\geqslant="3\AMSa3E

      \let\leq=\leqslant \let\le=\leqslant
      \let\geq=\geqslant \let\ge=\geqslant
    \fi
  \fi
\fi 

\ifnfsstwo
  \DeclareMathAlphabet{\mathbfit}{OT1}{cmr}{bx}{it}
  \SetMathAlphabet\mathbfit{bold}{OT1}{cmr}{bx}{it}
  \DeclareMathAlphabet{\mathbfss}{OT1}{cmss}{bx}{n}
  \SetMathAlphabet\mathbfss{bold}{OT1}{cmss}{bx}{n}
  \ifAMStwofonts
    \ifCUPmtlplainloaded \else
      \DeclareSymbolFont{UPM}{U}{eur}{m}{n}
      \SetSymbolFont{UPM}{bold}{U}{eur}{b}{n}
      \DeclareSymbolFont{AMSa}{U}{msa}{m}{n}
      \DeclareMathSymbol{\upi}{0}{UPM}{"19}
      \DeclareMathSymbol{\umu}{0}{UPM}{"16}
      \DeclareMathSymbol{\upartial}{0}{UPM}{"40}
      \DeclareMathSymbol{\leqslant}{3}{AMSa}{"36}
      \DeclareMathSymbol{\geqslant}{3}{AMSa}{"3E}

      \let\leq=\leqslant \let\le=\leqslant
      \let\geq=\geqslant \let\ge=\geqslant
    \fi
  \fi
\fi 

\ifCUPmtlplainloaded \else
  \ifAMStwofonts \else 
    \def\upi{\pi}
    \def\umu{\mu}
    \def\upartial{\partial}
  \fi
\fi

\title{Experimental Cosmic Statistics I~: Variance }
\author[S. Colombi et al.]{St\'ephane Colombi,$^{1}$ Istv\'an
Szapudi,$^{2,3}$ Adrian Jenkins,$^2$ and J\"org Colberg$^4$\\
$^1$Institut d'Astrophysique de Paris, CNRS, 98bis bd Arago,
F-75014 Paris, France\\
$^2$University of Durham, Department of Physics, South Road,
Durham, DH1 3LE, UK\\
$^3$Canadian Institute of Theoretical Astrophysics,
60 St George St, Toronto, Ontario, M5S 3H8 Canada (present address)\\
$^4$Max-Planck-Institut f\"ur Astrophysik, D-85740, Garching, Germany
}
\date{Submitted to MNRAS}
\pagerange{\pageref{firstpage}--\pageref{lastpage}}
\pubyear{1999}
\begin{document}
\voffset -0.5cm
\maketitle
\label{firstpage}
\begin{abstract}
Counts-in-cells are measured in the $\tau$CDM Virgo Hubble Volume
simulation. This  large $N$-body
experiment has $10^9$ particles in a cubic box of size 
$2000 \ h^{-1}$ Mpc. The unprecedented combination of 
size and resolution allows for the first time a realistic 
numerical analysis of the cosmic errors and cosmic
correlations of statistics
related to counts-in-cells measurements, such as the
probability distribution function $P_N$ itself, its factorial moments
$F_k$ and the related cumulants $\xiav$ and $S_N$'s.
These statistics are extracted from the whole
simulation cube, as well as from $4096$ 
sub-cubes of size $125\ h^{-1}$Mpc, 
each representing a virtual random realization of the local
universe. 

The measurements and their scatter over
the sub-volumes are compared to the theoretical predictions of 
Colombi, Bouchet \& Schaeffer (1995) for $P_0$, and of
Szapudi \& Colombi (1996, SC) and Szapudi, 
Colombi \& Bernardeau (1999a, SCB) for
the factorial moments and the cumulants.
The general behavior of experimental variance and 
cross-correlations as  functions of scale and order is well
described by theoretical predictions, with
a few percent accuracy in the weakly non-linear regime for
the cosmic error on factorial moments.
On highly non-linear scales, however, 
all variants of the hierarchical
model used by SC and SCB to describe clustering
appear to become increasingly approximate, which leads to a
slight overestimation of the error, by about a factor of
two in the worst case.  
Because of the needed supplementary perturbative approach,
the theory is less accurate for non-linear estimators,
such as cumulants, than for factorial
moments.

The cosmic bias is evaluated as well, and,
in agreement with SCB, is found to be insignificant compared to 
the cosmic variance in all regimes
investigated. 

While higher order statistics were previously evaluated in 
several simulations, this work presents text book quality
measurements of $S_N$'s, $3 \leq N \leq 10$, in an unprecedented
dynamic range of $0.05 \la \xiav \la 50$.  
In the weakly nonlinear regime 
the results confirm previous findings and
agree remarkably well with perturbation theory predictions 
including the one loop corrections based
on spherical collapse by Fosalba \& Gazta\~naga 1998. 
Extended perturbation theory is confirmed on all scales.

\end{abstract}
%
%
\section{Introduction}
%
%
Measurements of higher order statistics in galaxy catalogs
test theories of structure formation, the nature of the initial
fluctuations, and the processes of galaxy formation. The power of
such measurements to constrain theories, however, depends crucially on the
detailed understanding of the errors. 
Usually it is tacitly assumed that the underlying distribution of events
is Gaussian and thus
the term ``errors'' becomes synonymous with the ``variance''.
Knowledge of the variance is sufficient only when the
error distribution is Gaussian. 

For statistics related
to counts-in-cells a rigorous theory for the
cosmic errors was presented in a suite of papers by
Szapudi \& Colombi 1996, hereafter SC, Colombi, Szapudi \& Szalay
1998, and Szapudi, Colombi \& Bernardeau 1999a, hereafter SCB.
Nevertheless these calculations relied on approximations,
for which the domain of validity could not be checked extensively until
the arrival of the Virgo Hubble Volume Simulations.
Moreover, the regime where the underlying cosmic
distribution is Gaussian could not be examined previously.
This paper addresses the first problem by studying the statistical errors
and cross-correlations numerically, 
while a companion paper,
Szapudi et al.~(1999c, hereafter paper II), 
discusses the underlying distributions
of statistics in their full splendour.

Let us consider a statistic $A$ measured in a galaxy catalog 
of volume $V$. The corresponding indicator is denoted by $\tA$. 
In practice, only one sample of our local universe
is accessible. However, a frequentist numerical experiment
can be performed in a large numerical simulation
if a sufficient number $C_{\cal E}$ of galaxy catalogs  ${\cal E}_i$ can be
extracted from it. In each of them a value $\tA_i$, 
$1 \leq i \leq C_{\cal E}$, can be measured.

For any statistic $A$ the cosmic distribution function 
$\Upsilon(\tA)$ is the probability density of measuring
the value $\tA$ in a particular finite realization.
This distribution function can be approximately extracted
from the  $C_{\cal E}$ subsamples under the ergodic
hypothesis. For simplicity, we
dispense with the (logical) notation $\tilde\Upsilon$, and replace it
in what follows with $\Upsilon$. This 
expresses the fact that we do not wish
to enter one more level of complexity by considering
the ``error on the error'' problem (SC) in greater detail.
The smoothness and regularity of our
measurements suggest that the number of realizations, 
which represent a two orders of magnitude improvement over any previous
work, is large enough to provide an adequate determination of the
quantities measured. 

While in practice the function $\Upsilon(\tA)$ is the fundamental quantity
underlying all measurements, this paper concentrates
on its first two moments; paper II examines
its shape and skewness in detail.

In the following definitions, integrals are to be understood
as summations of the estimator over the distribution
function.
The first moment of $\Upsilon(\tA)$ 
is the spatial average
\begin{equation}
  \int \tA \Upsilon(\tA) d\tA=\langle \tA \rangle \equiv A,
\end{equation}
where it is assumed that the estimator $\tA$ is {\em unbiased}.
The bias is negligible compared to the relative cosmic error 
in most meaningful cases (SCB) as illustrated later by practical
examples. For completeness, however, the definition of the {\em cosmic bias} is
\begin{equation}
  b_A\equiv \frac{\langle \tA \rangle - A}{A}.
\end{equation}
The second (centered) moment of the cosmic distribution is called
the cosmic error, 
\begin{equation}
   \int (\tA-A)^2 \Upsilon(\tA)
   d\tA = \langle (\tA-A)^2 \rangle \equiv (\Delta A)^2.
\end{equation} 
For a biased statistic, the variance should be centered around
the biased average and not the true value. It can however be
shown formally (SCB) that the above definition is valid 
to second order in $\Delta A/A$ for any biased
statistic.\footnote{More precisely, to first order in $\langle
(\tx_i-x_i)(\tx_j-x_j) \rangle$ where $\tx_i$ denote the unbiased estimators
from which $\tA$ is constructed in a non-linear fashion.} 

Finally, the cosmic covariance can be defined
analogously to the variance as $\langle (\tA-A)(\tB-B) \rangle$.

The theoretical results for the errors and cross-correlations
are summarized below. If $v$ and $V$ are the cell and catalog volumes
respectively, 
the cosmic error can be approximately separated into three components
to leading order in $v/V$ (SC):
\begin{itemize}
\item The discreteness or shot noise error which is due to the finite number
of objects $N_{\rm obj}$ in the catalog. 
It increases towards small scales and with the order of
the statistics considered, but
becomes negligible when $N_{\rm obj}$ is very large.
\item The edge effect error is
due to the uneven weight given to galaxies near the edges of the survey
compared to those near the centre. It is especially significant on
large scales, comparable to the size of the catalog. 
\item The finite volume error is
due to fluctuations of the underlying density field
on scales larger than the characteristic size of the catalog.
\end{itemize}
The next to leading order correction in $v/V$
is proportional to the perimeter of the catalog $\partial V$.
At this level of accuracy there are also correlations  between
the three sources of error (e.g., Colombi et al.~1999, hereafter CCDFS).

Colombi, Bouchet \& Schaeffer (1995, hereafter CBS) 
investigated in detail the cosmic error on the void probability function. 
The groundwork for error calculations of statistics
related to counts-in-cells is based on SC where 
the cosmic error for factorial moments\footnote{e.g., Appendix A for 
definitions and notations.} was evaluated {\em analytically}.
SCB, extended the work of SC to cross-correlations, 
including perturbation theory
predictions (e.g., Bernardeau 1996). 
The cosmic errors, biases (see also Hui \& Gazta\~naga 1998, hereafter
HG) and covariances for cumulants$^{\textstyle \dagger}$ 
$\xiav$ and $S_N$ were calculated
as well. The main goal of this paper is to compare
the analytical predictions of CBS, SC and SCB to 
measurements made in the VIRGO $\tau$CDM Hubble Volume simulation.

The exhaustive nature of the comparison that follows warrants
the questions: is it meaningful to thrive for the detailed 
numerical understanding  of the theory? How much of it is 
practically useful? Can it accurately estimate the errors on 
measurements in future surveys? While some of these questions were addressed
in SCB, a brief account of supporting arguments is given next.

The analytics do take into account all possible 
theoretical errors, but systematics,
such as those resulting from cut out holes, incompleteness
from fiber separation, possible magnitude errors in the case of the
2dF, etc., could in principle corrupt
the theory and introduce biases. These effects might even require
detailed simulation of the survey. 
In the case of the UKST and Stromlo surveys such
simulations were performed and compared with the predictions: 
the spectacular agreement surprised even
the present authors (Hoyle, Szapudi, \& Baugh 1999).
Thus systematics do not dominate in all surveys;
for another example, where cut out holes found
to have insignificant effect on the cosmic probability
distribution of the two-point correlations 
function see Kerscher, Szapudi, \& Szalay (1999).

Moreover, the wide theoretical framework is flexible enough
to incorporate all systematics, which have the effect
of altering certain parameters, such as the factorial
moments. In such case any bias can be corrected for.

There might be unforeseen systematics which have such 
complicated non-linear effect that cannot even be modelled by appropriate
alteration of a set of parameters. While it would
be difficult to anticipate whether these could dominate for a particular
survey,  it is still instructive to investigate the potential
results in an ideal case, especially during design phase of the survey. 
Error calculations help optimizing geometry, sampling, and other parameters. 
During the design of the VIRMOS survey such considerations were
taken into account (Colombi et al.~1999). These calculations
as well as maximum likelihood analyses need to explore such a large
region in  parameter space, that they would typically be
impractical to carry out with simulations. 

In addition to applications to surveys, the theory
can be applied reliably to assess significance of measurements 
in simulations where multiple runs would
be too costly (e.g., Szapudi et al.~1999d).
All these present and potential future applications motivate 
the detailed investigations performed in this article.

The exposition is organized as follows. \S~2 describes the $N$-body 
data used for the purpose of our study.
\S~3 analyses the count-in-cells distribution 
function $P_N$, its cumulants $\xiav$ and $S_N$'s, 
and the scaling function of the void probability distribution
$\sigma\equiv -\ln(P_0)/F_1$. These quantities are measured
in the full simulation as well as in 
$C_{\cal E} = 4096$ subsamples. The accuracy of simulation is assessed
by comparing the measurements 
to the non-linear Ansatz of Hamilton et al.~(1991) 
improved by Peacock and Dodds (1996, hereafter PD), and to perturbation theory predictions 
(hereafter PT). The model of Fosalba \& Gazta\~naga (1998) and
extended perturbation theory (hereafter EPT, see
Colombi, Bernardeau, Bouchet \& Hernquist 1997) are considered as well.
\S~4 extends these investigations to the cosmic error and the variance of the cosmic
distribution function. A preliminary investigation
of the cross-correlations is done for factorial moments and cumulants.
The measurements are compared where possible to
the theoretical predictions of SC, SCB and CBS, including extended perturbation theory. 
Finally \S~5 recapitulates the results and discusses their implications.
In addition, Appendix~A gives a summary of the definitions and notations
used in this paper for counts-in-cells statistics. It will be useful for
the reader unfamiliar with these concepts.
%
%
\section{The $N$-body data}
%
%
The $\tau$CDM Hubble volume simulation (e.g., Evrard et al.~1999)
was carried out using a parallel 
${\rm P}^3{\rm M}$\ code described in MacFarland  et al.~(1998).
The code was run on 512 processors of the Cray T3E-600 at the Rechenzentrum
in Garching.  

Initial conditions were laid down by imposing perturbations on an
initially uniform state represented by a ``glass'' distribution of
particles generated by the method of White (1996). Because of the size
of the simulation, a glass file of $10^6$
particles was tiled 10 times in each direction. As
the initial glass file was created with periodic boundary conditions
tiling does not create any non-uniformities at the
interface between the tiles.  

A Gaussian random density field was set up by perturbing the positions of the
particles and assigning velocities to them according to the growing mode
linear theory solutions, using the algorithm described by Efstathiou
et al.~(1985). Individual modes were assigned random phases and the power
for each mode was selected at random from an exponential distribution
with mean power corresponding to the desired power spectrum $\langle |\delta_k^2| \rangle$. 
Unlike Efstathiou et al.~(1985), however, 
the initial velocities were set up exactly proportional to the initial
displacements, according to the Zel'dovich (1970) approximation. As
shown by Scoccimarro (1998) this leads to larger initial transients. To
compensate for this the simulation was started at a high redshift, $z=29$.

The cosmological model used for the simulation $\tau$CDM is described in
more detail in Jenkins et al.~(1998). 
The approximation to the linear CDM power spectrum
(Bond \& Efstathiou 1984) was used
\begin{equation}
 \langle |\delta_k^2| \rangle = {A k\over\bigg[1+\big[aq + (bq)^{3/2} 
+ (cq)^2\big]^{\nu}\bigg]^{2/\nu}}, 
\end{equation}
where $q = k/\Gamma$, $a = 6.4\ h^{-1}$ Mpc, $b = 3 \ h^{-1}$ Mpc, $c = 1.7 \ h^{-1}$ Mpc and
$\nu=1.13$. The value of $\Gamma$ was set equal
to $0.21$.  The normalization constant, $A$, is chosen by fixing the value of
$\sigma_8^2$ (the linear variance of the matter distribution in a sphere of
radius $8\ h^{-1}$ Mpc at $z=0$).
A value of $\sigma_8=0.6$ was motivated by estimates
based on cluster abundances (White, Efstathiou \& Frenk 1993; Eke, Cole \&
Frenk 1996).

 The simulation was integrated using a leapfrog scheme as described in 
Hockney \& Eastwood (1981), section 11-4-3. The simulation was completed
in 500 equal steps in time.  The softening used was 100 kpc/$h$ comoving Plummer 
equivalent - see Jenkins et al.~(1998).

%
%
\section{Counts-in-cells analysis~: the underlying statistics}
%
%

\begin{figure}
\centerline{\hbox{\psfig{figure=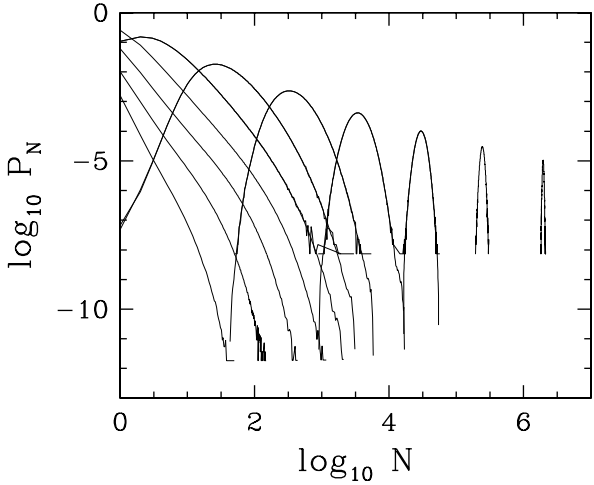,width=8cm}}}
\caption[]{The measured CPDF as a function of $N$. Various scales are
plotted as described in the text and in Table~\ref{table:table1}. The curves
shift to the right as $\ell$ increases.}
\label{fig:figure1}
\end{figure}

The count probability distribution function (CPDF) $P_N$ is defined as the
probability of finding $N$ objects in a cell of volume $v$ thrown at random in the
catalog.  CPDF was measured in the whole simulation ${\cal E}$ for cubic
cells of size $L_{\rm box}/512 \leq \ell \leq L_{\rm box}/8$, where $L_{\rm
box}=2000 \ h^{-1}$ Mpc is the size of the simulation cube (see Table~\ref{table:table1}).
Then the simulation cube was divided into 
$16^3$ contiguous cubic subsamples
${\cal E}_i$ of size $L=125 \ h^{-1}$
Mpc. $P_N$ was evaluated in each of these for $L/512 \leq \ell \leq
L/2$ (see Table~\ref{table:table1}). 
The successive convolution algorithm 
of Szapudi et al.~(1999d, hereafter SQSL) allowed the determination of the
CPDF on all scales simultaneously in only a few minutes of CPU on 
a workstation\footnote{This estimate does not include
the reading in of the file.} with $512^3$ sampling cells. 
The accuracy is thus $P_N \ge P_{\rm min,1}=1/512^3 
\simeq 7.45\times 10^{-9}$ for the measurement in ${\cal E}$ 
and for each individual ${\cal E}_i$; the accuracy increases by
averaging over all subsamples: 
$P_N \ge P_{\rm min,2}=1/(512\times16)^3 \simeq 1.82\times 10^{-12}$.
For $4 \la \ell  \la 63 \ h^{-1}$ Mpc
the measurements in ${\cal E}$ and
${\cal E}_i$ overlap (Table~\ref{table:table1}). This is illustrated by Fig.~\ref{fig:figure1},
displaying $P_N$ as a function of $N$: the figure
presents the CPDF extracted from both the full cube and 
averaged over all the sub-cubes. In the overlap region, the difference
can be detected as slight irregularities of the high-$N$ tail from the
full cube measurements.
The figure suggests that
at least on the smallest scales considered in
${\cal E}$ (or each ${\cal E}_i$), our sampling is probably insufficient
by the standards of SC. 
However, this does not affect significantly the
calculations as indicated by the agreement of the moments
measured in ${\cal E}$ and those
calculated from averages obtained from the subsamples. 
Therefore measurement errors will be neglected in what follows,
i.e. {\em infinite sampling} is assumed.
Note that this ideal can be achieved in practice for two-dimensional
and small three-dimensional catalogs via
the algorithm of Szapudi (1998), however, the present 
simulation is too large for this method.

The smallest scale considered is only 
$2.4$ times larger than the softening length
$\lambda_{\epsilon}= 100 \ h^{-1}$ kpc. 
As discussed extensively in Colombi, Bouchet \& Hernquist (1996), contamination by softening 
restricts the validity of the simulation on small scales.
For spherical cells of radius $R$, at least $R\ga
4\lambda_{\epsilon}$ should hold.
For the cubic cells of the present simulation this condition
translates to $\ell \ga 6.5
\lambda_{\epsilon} \simeq 0.65\ h^{-1}$ Mpc. Thus
the two smallest cell sizes  i.e. the two leftmost points 
could be contaminated by softening, a fact that should be borne in mind,
especially when comparing with theoretical calculations which
employ models motivated by dynamics.
On the other hand, for statistical purposes
the dynamics can be ignored and the simulation can be
regarded as a set with prescribed statistics.
Then the possible contamination is irrelevant at the level of the  approximations
taken in the next sections. 

Another possible source of contamination could be in principle the
anticorrelation introduced by the glass initial positions. The effect
of this is, however,  extremely small as evidenced by
the measurement of ${\bar \xi}$ shown below.
\begin{table*}
\caption{The scales for which we measured the CPDF}
\begin{tabular} {l|ccccccccccc}
\hline 
$\ell (h^{-1}\ {\rm Mpc})$ & $0.24$ & $0.49$ & $0.98$ & $1.95$ &
                             $3.91$ & $7.8$  & $15.6$ & $31.3$ & 
                             $62.5$ & $125$  & $250$         \\ \hline
${\cal E}$                 &        &        &        &        &
                            $\surd$ &$\surd$ &$\surd$ &$\surd$ &
                            $\surd$ &$\surd$ &$\surd$       \\
${\cal E}_i$               &$\surd$ &$\surd$ &$\surd$ &$\surd$ &
                            $\surd$ &$\surd$ &$\surd$ &$\surd$ &
                            $\surd$ &        &              \\ \hline
\end{tabular}
\label{table:table1}
\end{table*}

\begin{figure}
\centerline{\hbox{\psfig{figure=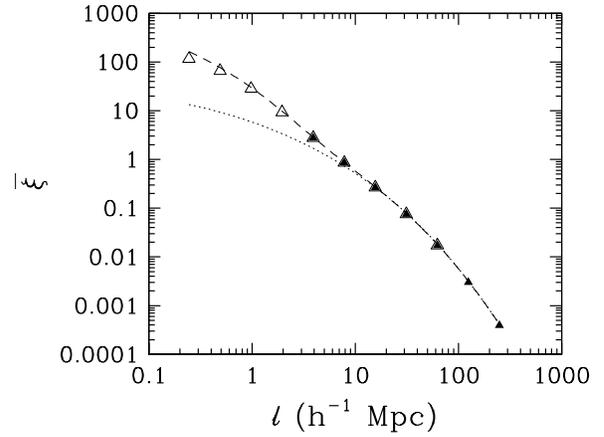,bbllx=25pt,bblly=207pt,bburx=571pt,bbury=606pt,width=8cm}}}
\caption[]{The averaged two-point correlation function ${\bar \xi}$ as
a function of scale. It is compared with linear theory (dots) and with
the non-linear Ansatz of Hamilton et al.~(1991) with the recipe of
Peacock \& Dodds (1996) (dashes). The open symbols correspond to the
${\xiav}$ obtained from the CPDF averaged over all the subsamples
${\cal E}_i$ and the filled symbols to the measurement in ${\cal E}$.}
\label{fig:figure2}
\end{figure}

Figure~\ref{fig:figure2} displays the average correlation
function ${\bar \xi}$  as a function of scale. By definition
\begin{equation}
  \xiav \equiv \frac{1}{v^2} \int_v d^3r_1 d^3r_2 \xi(|r_1 - r_2|),
\end{equation}
where $\xi(r)$ is the two-point correlation function.
In practice, it is obtained as the variance of the counts-in-cells, 
corrected for discreteness effects automatically via the
use of factorial moments (e.g., see SQSL and Appendix A for the
detailed description of the method used in this paper to
obtain the cumulants including the variance from counts-in-cells). 
The measured $\xiav$ is compared with
linear theory (dots) and with the non-linear Ansatz of Hamilton et
al.~(1991) improved by PD (dashes). As expected,
the agreement with linear theory in the regime $\xiav \la 1$ 
is excellent, even on the largest
scales where the  anticorrelations introduced by the glass 
initial condition could cause contamination.
The two leftmost points are slightly below the dashes, 
because of softening effects as discussed above, otherwise the
results are in perfect accord with theory.

\begin{figure*}
\centerline{\hbox{\psfig{figure=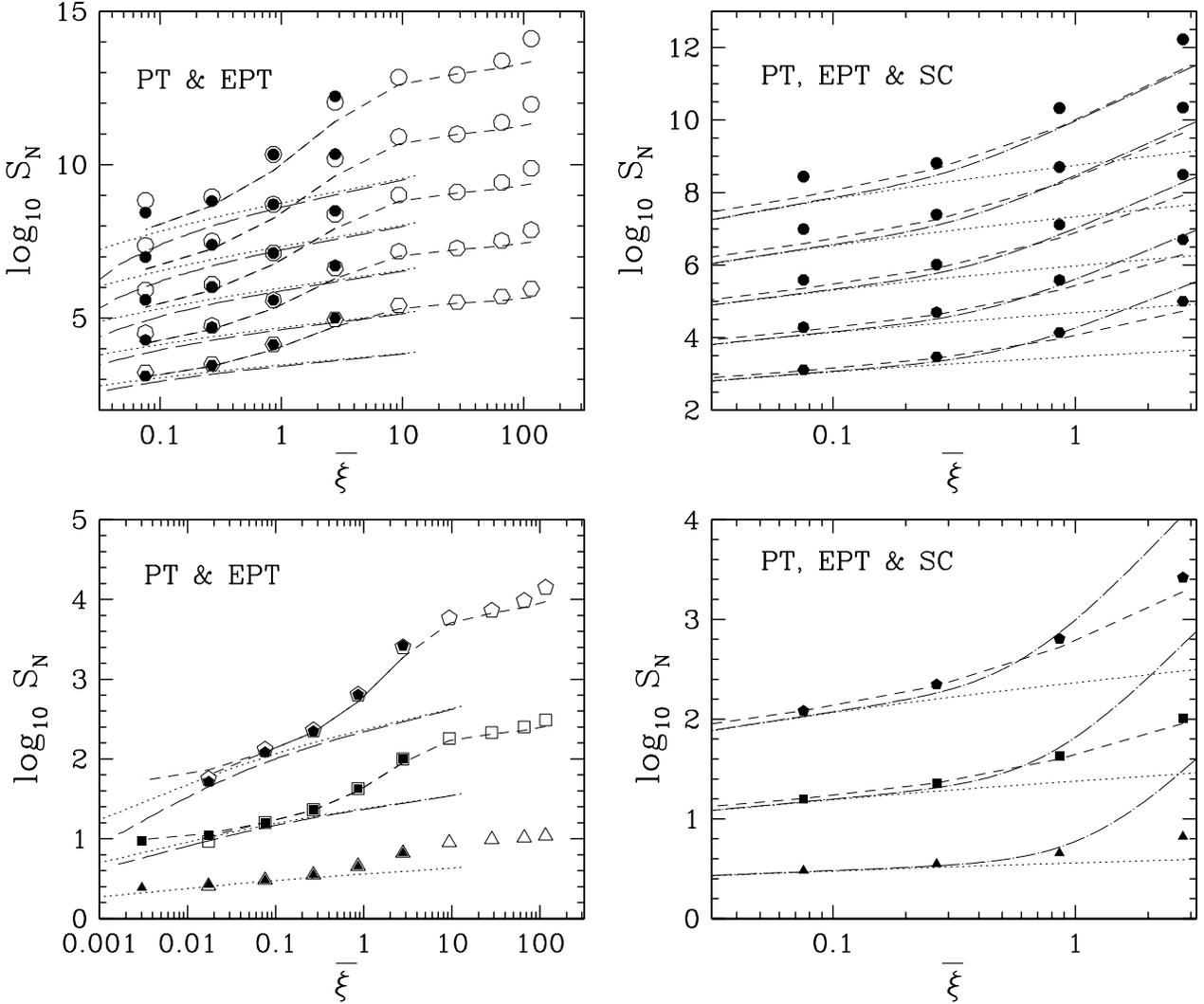,bbllx=62pt,bblly=297pt,bburx=512pt,bbury=675pt,width=17cm}}}
\caption[]{The cumulants $S_N\equiv {\bar \xi}_N/{\bar \xi}^{N-1}$ as
functions of ${\xiav}$ compared to various theoretical models.
The left-hand panels show the full dynamic range, while the right-hand
ones concentrate on the transition to the non-linear regime.
The models considered are perturbation theory (dots on all panels and
long dashes on left panels), extended perturbation theory (short dashes), 
and one loop perturbation theory based on the spherical model (dots-long dashes
on right panels). The upper and the
lower panels give $S_N$ for $6 \leq N \leq 10$ and $3 \leq N \leq 5$
respectively (the value of $S_N$ increases with order $N$). 
The convention for the symbols is the same as in 
Fig.~\ref{fig:figure2}. Note that the right-hand panels show only
the measurements in the full simulation ${\cal E}$.}
\label{fig:figure3}
\end{figure*}

Figure~\ref{fig:figure3} plots the extracted cumulants $S_N$'s against
$\xiav$. They are compared with predictions of various models, including 
perturbation theory (PT, dots). By definition
(e.g., Balian \& Schaeffer 1989a)
\begin{equation}
   S_N = N^{N-2} Q_N \equiv {\bar \xi}_N/\xiav^{N-1},
\end{equation}
where ${\bar \xi}_N$ is the $N$-point correlation 
function averaged over a cell:
\begin{equation}
   {\bar \xi}_N = \frac{1}{v^N} \int_{v} d^3r_1 \cdots d^3r_N \xi_N(r_1,\cdots,r_N).
\end{equation}
Perturbation theory predictions have been calculated for spherical cells
by Juszkiewicz, Bouchet \& Colombi (1993) for $S_3$ and extended to
arbitrary order by Bernardeau (1994):
\begin{equation}
  S_N(\ell)=f_N(\gamma_1,\cdots,\gamma_{N-2}),
  \label{eq:snef}
\end{equation}
\begin{equation} 
  \gamma_i\equiv \frac{d^i\log \xiav}{(d\log \ell)^i}.
\end{equation}
For example 
\begin{equation}
  S_3=\frac{34}{7}+\gamma_1,
\end{equation}
\begin{equation}
  S_4=\frac{60712}{1323}+\frac{62}{3}\gamma_1+\frac{7}{3} \gamma_1^2
  -\frac{2}{3} \gamma_2.
\end{equation}
The dots on Fig.~\ref{fig:figure3} assume $\gamma_i=0$, $i \geq 2$. 
While this is incorrect in
principle for a scale dependent spectrum such as
$\tau$CDM,  the long dashes on the left-hand panels 
prove that the contribution of  
$\gamma_2$ is insignificant.
Higher order $\gamma_i$ terms, as discussed also by Baugh,
Gazta\~naga \& Efstathiou (1995), have an even smaller effect and can be
rightly neglected. 

PT predictions are accurately fulfilled
in the weakly non-linear regime. This
confirms again numerous earlier works (see, e.g. 
Juszkiewicz, Bouchet \& Colombi 1993; Bernardeau 1994; 
Juszkiewicz et al.~1995; Gazta\~naga \& Baugh 1995; Baugh, 
Gazta\~naga \& Efstathiou 1995; 
SQSL). In
fact the textbook quality agreement with PT demonstrates the
accuracy of the $\tau$CDM Hubble 
Volume simulation.

The dashes give the predictions obtained from extended
perturbation theory (EPT, Colombi et al.~1997; see also 
Szapudi, Meiksin \& Nichol 1996 for EPT applied to
galaxy data, and Scoccimarro \& Frieman, 1998 for ``hyperextended'' 
perturbation theory). EPT assumes that the same forms of the higher
order moments are preserved in the highly non-linear regime.
There $\gamma_1$ above is simply an adjustable parameter without any
particular meaning, i.e.
\begin{equation}
  \gamma_{1,{\rm eff}}=\gamma_1(S_3)=S_3-\frac{34}{7},
  \label{eq:gammaeff}
\end{equation}
where $S_3$ is the measured one.
With this value of $\gamma_1$ the $S_N$'s, $N
\geq 4$, can be computed  using equation (\ref{eq:snef}) 
(with $\gamma_i=0$, $i\geq 2$). 
The dashed curves matches the measurements
quite well even in the highly non-linear regime thereby
reconfirming the efficiency of EPT (see also SQSL). 
The agreement is not expected to be absolutely perfect from this
Ansatz: on Fig.~\ref{fig:figure3}, EPT tends to underestimate slightly
the measured values  of $S_N$ when $1 \la \xiav \la 10$.

The dynamic range in the upper left panel of 
Fig.~\ref{fig:figure3} is narrower than in the lower left panel:
on large scales the agreement between PT and
measurement becomes less accurate for the $S_N$'s,
especially if $N$ is large. 
This might be related to transients
due to the initial setup of the particles on 
a glass perturbed by using the Zel'dovich approximation. 
On the one hand, the transients
related to pure Zel'dovich should decrease the value of the $S_N$'s (e.g.,
Juszkiewicz et al.~1993 and Scoccimarro 1998) while,  on the other hand,
the anticorrelations due to the glass could have the opposite 
effect by decreasing $\xiav^{N-1}$ more than $\xiav_N$. Although
this problem was not examined in detail, the
glass contamination on $\xiav$ appears to be inconsequential.
Alternatively, finite volume effects can 
degrade the high-$N$ tail of the CPDF (e.g., Colombi, Bouchet \&
Schaeffer 1994; CBS; Colombi et al.~1996). In addition, it is worth reemphasising
that two rightmost points are prone to errors caused
by softening as discussed earlier.

The right-hand panels of Fig.~\ref{fig:figure3} zoom in on
the transition between the weakly and highly nonlinear regime. 
For comparison, PT (with $\gamma_i=0$, $i\geq 2$, dots), 
EPT (dashes), and the one loop perturbation theory
of Fosalba \& Gazta\~naga (1998)
(dots-long dashes) are displayed. The last model yields agreement
with the extracted values of $S_N$ for $\xiav \la 1$, or even larger
when the order $N$ is high enough (see upper right panel). 
This affirms the success of one-loop perturbation theory
(see also Lokas et al.~1996; Scoccimarro et al.~1998).
Interestingly, EPT produces almost identical results to the spherical model 
when $\xiav \la 1$. 

\begin{figure}
\centerline{\hbox{\psfig{figure=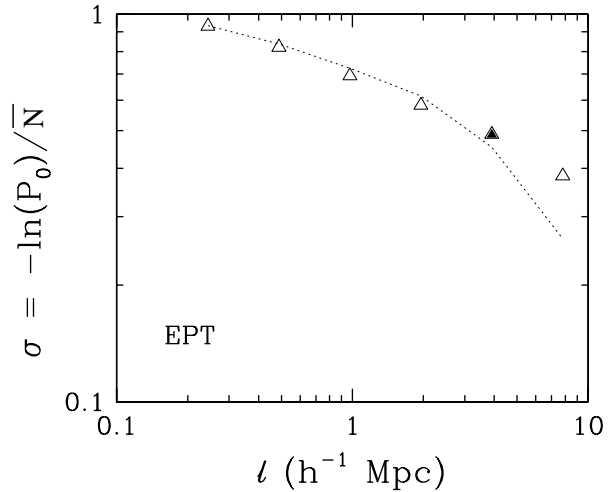,bbllx=35pt,bblly=205pt,bburx=557pt,bbury=641pt,width=8cm}}}
\caption[]{The scaling function $\sigma\equiv -\ln(P_0)/{\bar N}$,
compared with extended perturbation theory (dots).  
The convention for the symbols is the same as in 
Fig.~\ref{fig:figure2}. Note that on the largest scales
we measure $P_0=0$, and thus no
points are plotted.
For the direct measurement in ${\cal E}$ there is no empty
cell with $\ell=7.8 \ h^{-1}$ Mpc because of our insufficient sampling.}
\label{fig:figure3bis}
\end{figure}

Finally, figure~\ref{fig:figure3bis} shows 
$\sigma=-\ln(P_0)/{\bar N}$ as a function of scale, compared with EPT
predictions. By definition
(White 1979; Balian \& Schaeffer 1989a; see also Appendix A)
\begin{equation}
  \sigma=\sum_{N=1}^{\infty} (-1)^{N-1}\frac{S_N}{N!}
  \left({\bar N}\,\xiav \right)^{N-1},
  \label{eq:sigsig}
\end{equation}
where ${\bar N}$ is the average count in a cell.
This function is thus sensitive to low order statistics when
$N_{\rm c} \equiv {\bar N} \,\xiav \ll 1$, and to high order statistics
when $N_{\rm c} \gg 1$. According to Fig.~\ref{fig:figure3bis}, 
EPT is an accurate Ansatz on small scales where $\sigma$ is
close to unity and is dominated by low order $S_N$. It is a less
precise approximation on the largest scales probed, as expected.
Indeed, the rightmost point of Fig.~\ref{fig:figure3bis}
corresponds to where $\xiav\simeq 1$ in
Fig.~\ref{fig:figure3}. There EPT increasingly
underestimates the $S_N$'s when $N$ is high.
Note the remarkable power-law behavior of 
$\sigma \propto \ell^{-D_0}$, $D_0 \simeq 0.25$,
in agreement with the predictions of the
scaling model of Balian \& Schaeffer (1989a). 
This reflects a non-trivial (multi)fractal
particle distribution (Balian \& Schaeffer 1989b) with a Hausdorff 
dimension $D_0$. Such behavior
was found in a standard CDM model by Bouchet, Schaeffer \&
Davis (1991). Subsequently, the fractal distribution 
with $D_0 \simeq 0.5$ was established by
Colombi, Bouchet \& Schaeffer (1992). 
%
%
%
%
\section{The Cosmic Error}
%
%

  In the previous section we demonstrated that good agreement was obtained comparing
measurements made on the $\tau$CDM Hubble Volume dataset with previous
work regarding higher order clustering statistics. Having established
the accuracy of the dataset this section concentrates on the
the determination of cosmic errors
and their comparison to the available theoretical predictions, where
possible.  In \S~4.1 we summarise analytic calculations of the cosmic
errors and their cross-correlations. From this follows a systematic
study of the experimental cosmic error of low-order statistics,
i.e.~factorial moments $F_k$, $1 \leq k \leq 4$ (\S~4.2), and
cumulants $\xiav$, $S_3$ and $S_4$ (\S~4.3) together with a 
thorough comparison with the theoretical predictions. Also in
\S~4.3 we discuss the cosmic bias of the cumulants. Then the
void probability and its scaling function $\sigma$ are explored
(\S~4.4) followed by the cosmic error on the CPDF itself (\S~4.5).
Finally, in \S~4.6, there is a preliminary investigation of the cosmic 
cross-correlations of factorial moments and cumulants.

In all subsequent figures, except for
the cross-correlations, there are errorbars plotted on 
the symbols corresponding to measurements due to the 
finite number of realizations $C_{\cal E}=4096$. 
These measurement errors, proportional to $1/\sqrt{C_{\cal E}}$ (SC), are
negligible for our simulation, and the errorbars 
are smaller than the size of the symbols in most cases.
As discussed in the Introduction, we neglect the cosmic error on
the determination of the cosmic error (which is due to the finite size
of the Hubble Volume itself) because in practice it is insignificant.
%
%
\subsection{Cosmic Error: Theoretical Predictions}
%
Before making any comparison with the analytic predictions,
we outline the main ideas in CBS, SC, and SCB -- more details
can be found in these papers.
Spherical cells  of radius $\ell$ are assumed throughout for simplicity.

The bivariate CPDF $P_{N,M}(\ell,r)$ is the probability
of finding $N$ and $M$ points
in two cells of size $\ell$ at distance $r=|r_1-r_2|$ from each other. 
According to SC the cosmic error
is computed via a double integral of  $P_{N,M}(\ell,r)$ over
$r_1$, and $r_2$, conveniently split according to 
whether the cells overlap or not:
\begin{description}
\item[(i) {\em overlapping cells} ($r \la 2\ell$):] give rise to
the discreteness and edge effect errors (see Introduction).
The locally Poissonian assumption (CBS, SC) enables the
approximate representation of the generating function $P(x,y)$ 
for overlapping cells by using only the monovariate generating function 
$P(x)$,  i.e. the calculation depends on $\xiav$, $S_N$, $N\geq 3$ and the
average count ${\bar N}$.
\item[(ii) {\em disjoint cells} ($r \ga 2\ell$):] generate
the finite volume error (see Introduction).
To simplify the writing of
$P_{N,M}(\ell,r)$, the distance $r$ is assumed to be 
large enough compared to the cell
size such that the bivariate CPDF can be Taylor expanded (to first order)
in terms of $\xi(r)/\xiav$. This approximation is surprisingly accurate
even when the cells touch each other (Szapudi, Szalay, \& Bosch\'an
1992, Bernardeau 1996, hereafter B96).
Three models are used: two particular but still
quite general forms of the hierarchical model, SS and BeS, introduced
by Szapudi \& Szalay (1993a, hereafter SSa, 1993b) and
by Bernardeau \& Schaeffer (1992), respectively,  and perturbation theory,
hereafter PT (B96).  See SC and SCB for more details.
The former two models depend only on monovariate statistics,
i.e. on $\xiav$ and $S_N$, $N \ge 3$ and ${\bar N}$. 
PT on the other hand is expressed in terms of $\gamma_i$,
$\xiav$ and ${\bar N}$ (B96). 
In principle, PT is accurate only in the weakly non-linear regime, for which it
was originally designed, but it can be extended to the nonlinear
regime as well: for monovariate distributions, EPT was proposed
by Colombi et al. (1997), as discussed and tested versus
measurements in \S~3. This Ansatz can actually  be 
naturally generalized to the bivariate
CPDF (Szapudi \& Szalay 1997, SCB). Our version, denoted by E$^2$PT,
takes the measured (non-linear) value for $\xiav$,
$\gamma_{1,{\rm eff}}$ from equation~(\ref{eq:gammaeff}) and it
assumes, as EPT, $\gamma_i =0$ for $i \geq 2$.
\end{description}

Except for the error on the void
probability and its scaling function $\sigma$ detailed
in CBS, the theoretical results shown in this section were computed to 
leading order in $v/V$, where $v$ is the cell volume and $V=L^3$ 
is the sample volume. 

The calculation of the error on a statistics of order $k$ depends on 
${\bar N} \equiv F_1$,  $\xiav$, $\xiav({\hat L})$, the average
of the correlation function over the survey (see below),
and $S_N$, $3 \leq N \leq 2k$. PT is determined by
$\gamma_i$, $i \leq 2k-2$ (\S~3) and E$^2$PT by $\gamma_{1,{\rm eff}}$
as explained above. In all cases, we use the
measured value of ${\bar N}$. Other parameters are chosen as follows:
\begin{enumerate}
\item[(a) {\em PT:}] linear theory is employed to compute $\xiav$
and $\xiav({\hat L})$ [the catalog is assumed to be spherical to
simplify the calculation of integral (\ref{eq:xihat}) below] while
higher order statistics are evaluated according to eq.~(\ref{eq:snef})
with $\gamma_i=0$, $i\geq 2$.
\item[(b) {\em Other models:}]: the experimental $\xiav$ is used
(open symbols on Fig.~\ref{fig:figure2}). The quantity $\xiav({\hat L})$ is
computed numerically with the non-linear Ansatz of PD discussed in \S~3
(assuming that the catalog is spherical).
For the $S_N$'s, the measurements (open symbols on left panels of
Fig.~\ref{fig:figure3}) are used for $\ell
\le 15 \ h^{-1}$ Mpc. On larger scales, EPT is more appropriate
to determine $S_N$, $N \geq 4$: the increasing inaccuracy 
of the $S_N$'s on large scales and
for large $N$ require this procedure. It is 
justified all the more since when $\xiav \la 0.27$ EPT matches quite well 
to the  PT predictions (see Fig.~\ref{fig:figure3}).

There is a subtlety worth mentioning which concerns
the finite volume error, proportional to the
integral
\begin{equation}
  {\xiav({\hat L})}=\frac{1}{\hat V} \int_{r_{12} \geq 2\ell} d^3r_1 d^3r_2
  \xi(|r_1-r_2|).
\end{equation}
To leading order in $v/V$, this integral reads (CCDFS)
\begin{equation}
  {\xiav({\hat L})}=
  \xiav_0({\hat L})-\frac{8 v}{\hat V} \xiav_1(2\ell),
  \label{eq:xil}
\end{equation}
with
\begin{equation}
  \xiav_0({\hat L})=\frac{1}{{\hat V}^2} \int_{r_1,r_2 \in {\hat V}}
  d^3r_1 d^3r_2 \xi(|r_1-r_2|),
  \label{eq:xihat}
\end{equation}
\begin{equation}
  \xiav_1(\ell)\equiv \frac{1}{v} \int_{r \leq \ell} 4\pi r^2
  \xi(r)dr.
\end{equation}
In the above equations, ${\hat V}$ 
corresponds to the volume covered
by cells of volume $v$ included in the catalog.

The next to leading order correction, $\xiav_1$, 
can be identified as a negligible correction to the 
edge effects for most practical purposes. 
Although it did not make a significant
difference, we included this correction nonetheless.

\end{enumerate}
\subsection{Cosmic Error: Factorial Moments}
\begin{figure}
\centerline{\hbox{\psfig{figure=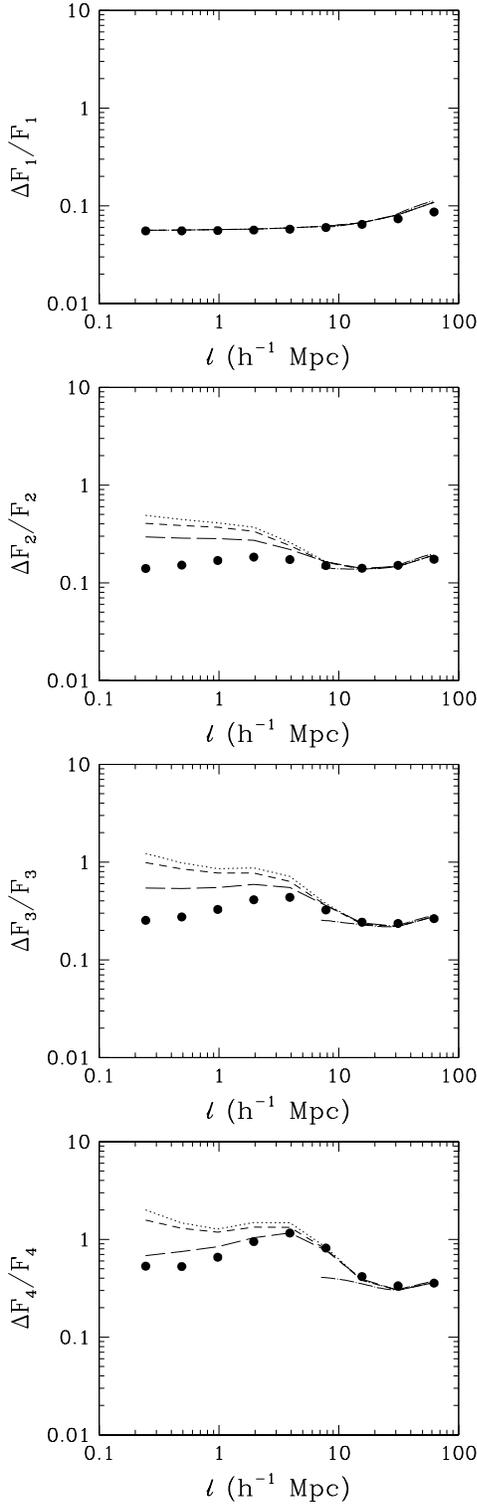,bbllx=180pt,bblly=84pt,bburx=382pt,bbury=708pt,width=6.5cm}}}
\caption[]{The cosmic error $\Delta F_k/F_k$ as a function of
scale. Each panel corresponds to a value of $k$. The dots, dashes, long
dashes, dot-long dashes correspond respectively to the SS, BeS, E$^2$PT and PT 
models. PT is shown only in its expected range of validity, $\ell \ga
\ell_0$, where $\ell_0$ is the correlation length defined by
$\xiav(\ell_0)\equiv 1$.
For $k=1$, all the models give the same result. As discussed in the
beginning of \S~4, there are errorbars
due to the finite number of realizations $C_{\cal E}=4096$, but they
are so small that they do not show.}
\label{fig:figure4}
\end{figure}

%
Figure~\ref{fig:figure4} presents the cosmic error measured for the
factorial moments $F_k$, $1 \leq k \leq 4$. 
By definition 
\begin{equation}
   F_k \equiv \langle (N)_k \rangle \equiv
   \langle N(N-1)\cdots (N-k+1) \rangle=\sum_N (N)_k P_N.
\end{equation}
The factorial moments directly estimate the moments of the 
underlying continuous density field: $F_k = {\bar N}^k 
\langle \rho^k \rangle$ where ${\bar N}=F_1$ is the average count (e.g., SSa).
On Fig.~\ref{fig:figure4}, the dotted, dash, long dash and 
dotted-long dash curves correspond to SS,
BeS, E$^2$PT and PT. 

\begin{figure}
\centerline{\hbox{\psfig{figure=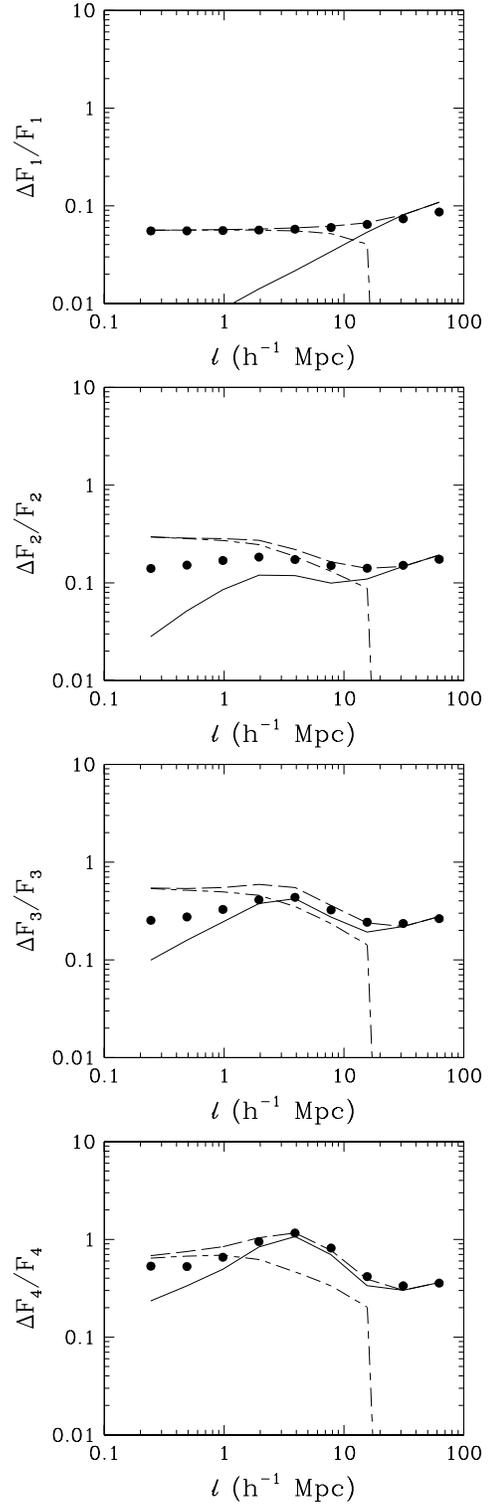,bbllx=180pt,bblly=84pt,bburx=382pt,bbury=708pt,width=6.5cm}}}
\caption[]{Same as in Fig.~\ref{fig:figure4}, but now the long
dashed, dashed-long dashed and solid curve correspond respectively 
to the E$^2$PT model, the finite volume contribution, and the edge +
discreteness contribution. Note the sudden cut-off at large scales for
the finite volume error, in agreement with eq.~(\ref{eq:xil}). Without
the $8(v/V) \xi_1(2\ell)$ correction, the cut-off would not show
up, but this would not significantly change the total error.}
\label{fig:figure5}
\end{figure}

All the models converge and agree quite well with the measurements 
on large scales $\ell \ga \ell_0\simeq
7.1 \ h^{-1}$ Mpc, as expected, since PT predictions should be valid.
In contrast, on small scales $\ell < \ell_0$ the models
overestimate slightly the numerically obtained error,
E$^2$PT being the most accurate.
It is worth remembering
that the leftmost two points may be contaminated by
smoothing effects and should not be over-interpreted.
Nevertheless, the decrease of precision on small scales suggests that 
our assumptions (i) or (ii) in \S~4.1 are becoming more and
more approximate
in the non-linear regime, i.e. either the local Poisson
assumption or the particular hierarchical
decompositions loose their accuracy. 
To test this idea
the contribution of overlapping cells (edge $+$
discreteness effects) were separated from the contribution of 
disjoint cells (finite
volume effects), as shown respectively
as solid and dash-long dash curves
on Figure~\ref{fig:figure5},
which concentrates on  E$^2$PT (long dashes). 
Note that the solid curve represents the SS and BeS models as well.
Finite volume
effects appear to dominate on small scales because our subsamples are 
dense enough to suppress discreteness error as expected (SC). 
This pinpoints assumption (ii) as the source of inaccuracy. Note that
naively one would suspect additional loss of precision 
in the Taylor expansion of the bivariate
CPDF. However, the finite volume error is a double integral over all the
cells included in the catalog and separated by more than
$2\ell$. The contribution of close cells is small, especially
when $\ell/L$ is small. Thus E$^2$PT itself appears to break down in the
non-linear regime (SS and BeS are even less accurate), at least for the
particular experiment we are analysing. Despite that EPT itself fares
quite well (Fig.~\ref{fig:figure3}), its
simplest natural extension to bivariate distributions, E$^2$PT, 
is less accurate, as noticed earlier by Szapudi \& Szalay (1997) 
in connection with the cumulant correlators of the APM galaxy catalog. 
However, the accuracy of the calculation based on  E$^2$PT
should be adequate for most practical uses, and future work on 
the representation of the bivariate distribution in the highly non-linear
regime will result in increased precision.

The solid curves in Fig.~\ref{fig:figure5} represent the main contribution
of the cosmic error on large scales.  Here, as expected (SC),
the cosmic error is  dominated by edge effects.
Despite the fact that 
theoretical predictions were determined to leading order in $v/V$
and the largest scale considered is $\ell=L/2$, i.e.~$v/V=1/8$,
the agreement between theory and measurement is surprisingly good. 
CCDFS have computed the next to leading 
order contribution proportional to the perimeter $\partial V$
of the survey.  With this correction, which increases the cosmic error
especially on the largest scales, next to leading order theory would be
inferior to the leading order one. The reason is that 
the calculation of CCDFS assumes a perimetric curvature radius 
much larger than the cell size. 
This assumption, which is useful for deep galaxy
surveys with small sky coverage,  obviously fails for a compact catalogue
such as this one where the cell size $\ell$ becomes comparable to
$L$.

\subsection{Cosmic Error and Cosmic Bias: Variance and Cumulants}
So far only the full moments $F_k$ have been examined. 
The cumulants $\xiav$ and $S_N$, however, are the more
physically motivated quantities.
But the statistics of these is complicated
by the fact that they are ratios. For example
(see Appendix A)
\begin{equation}
  \xiav=F_2/F_1^2-1.
\end{equation}
As is well known in statistics (e.g., HG, SCB) 
$\langle A/B \rangle \neq \langle A \rangle/\langle B
\rangle$. In other words, the estimator
\begin{equation}
  {\tilde{\xiav}}=\tF_2/\tF_1^2-1
\end{equation} 
is biased. Note that this is a general feature for any statistic
constructed from unbiased estimators in a non-linear fashion
(e.g. SCB). However, SCB showed theoretically 
that the {\em cosmic bias} defined in the Introduction, 
given here by 
\begin{equation}
   b_{\xiav}\equiv (\langle {\tilde{\xiav}} \rangle - \xiav)/\xiav,
\end{equation}
is of same order of $(\Delta {\xiav}/{\xiav})^2$ in the regime
$\Delta {\xiav}/{\xiav} \ll 1$. Similar 
reasoning applies to the $S_N$'s. Thus leading order
theoretical calculations neglect the bias. 
This can be done safely in the domain of validity 
of the perturbative approach used to expand a non-linear combination
of biased estimators. A reasonable criterion
proposed by SCB for this domain is
that the cosmic bias be small compared to
the relative cosmic error which itself should be small compared 
to unity. For an arbitrary (possibly biased) statistic $A$ this
reads
\begin{equation}
  b_A \ll \Delta A/A \ll 1.
 \label{eq:criterion}
\end{equation}

\begin{figure*}
\centerline{
\hbox{\psfig{figure=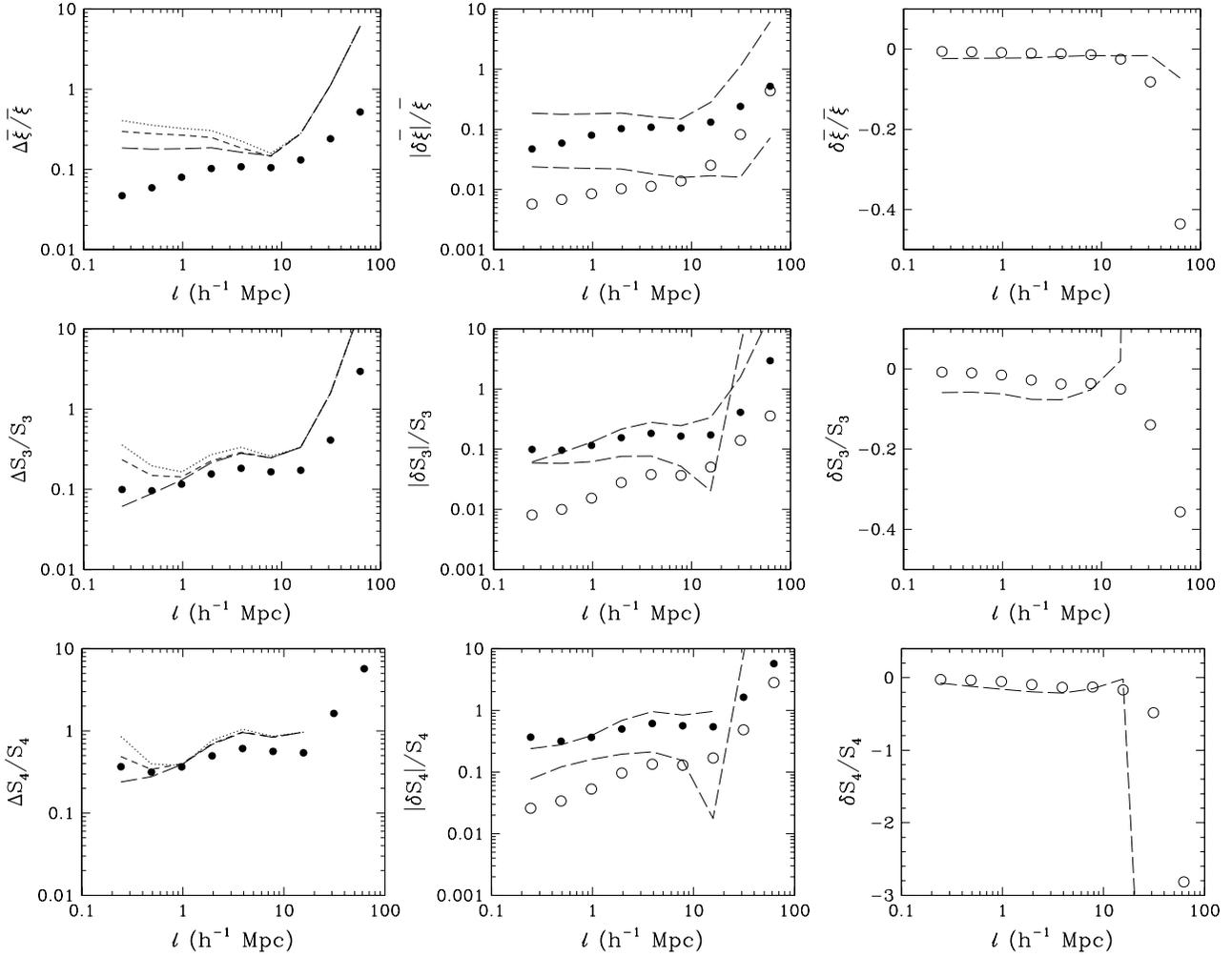,bbllx=83pt,bblly=311pt,bburx=537pt,bbury=673pt,width=17cm}}}
\caption[]{Same as Fig.~\ref{fig:figure4} for the
average correlation function (top row of panels), and the cumulants $S_3$
(middle row of panels) and $S_4$ (lower row of panels). 
The cosmic bias is plotted both in logarithmic coordinates 
(middle column of panels) and linear
coordinates (right column of panels). 
The filled and the open symbols correspond to
the cosmic error and the cosmic bias respectively. 
The theory breaks down on large scales 
for $S_4$ shown in the bottom left-hand panel. In this regime, the leading order
calculation gives negative $(\Delta S_4/S_4)^2$ (see SCB). 
The theory result for the cosmic bias is shown for the E$^2$PT model only. 
In the middle column panels, there
are two long-dashed curves: each one of which should be compared with
the closest symbols overall, corresponding either to the cosmic error 
(filled) or the cosmic bias (open).}
\label{fig:figure6}
\end{figure*}

Left panels of Figure~\ref{fig:figure6} are analogous
to Fig.~\ref{fig:figure4} 
and show the measured cosmic error 
as a function of scale for the biased estimators of $\xiav$,
$S_3$ and $S_4$. The middle panels show the absolute value of the cosmic
bias (open symbols) compared to the cosmic error (filled symbols). 
For additional clarity, the cosmic bias is plotted in linear coordinates 
as well in the right-hand panels.

It is interesting first to compare the cosmic error for factorial
moments and cumulants of same order.
The discreteness error is negligible
for the scaling regime and the statistics considered here.
The cumulants fare better/worse than the factorial moments
in the non-linear/weakly nonlinear regimes, respectively.
The finite volume error, dominating on small scales,
is the limiting factor for factorial moments, while
the edge effect error, dominating
on large scales, drives the errors of
the cumulants. 
This is in full accord with the 
predictions of SCB which can be consulted for more details.

The theoretical models on Fig.~\ref{fig:figure6} use the analytic calculations
of SCB and are computed analogously to Fig.~\ref{fig:figure4}, as
explained in \S~4.1.
E$^2$PT only is presented in the  middle and right-hand panels.
Again, it is worth remembering that the leftmost points are
dangerously close to the limit of possible contamination from
artificial smoothing effects introduced by the force softening.

For the variance $\xiav$, the theory systematically
overestimates the errors and the cosmic bias, except for the latter on 
large scales. This is not at all unexpected in light of the previous findings
on small scales, where the three models SS, BeS and E$^2$PT loose precision.
In the weakly non-linear regime, $\ell > \ell_0=7.1\ h^{-1}$ Mpc,
where perturbation theory is valid, this is 
somewhat disappointing. 
However, the dynamic range is limited by the criterion (\ref{eq:criterion}),
which is hardly, if at all, fulfilled here.
Hence the leading order perturbative approach
is likely to be insufficient. 

For higher order statistics $S_3$ and $S_4$,
the theory again tends to overestimate the amplitude 
of the measured cosmic bias on small scales.
On large scales, where the predicted 
$|b_{S_k}|$ presents a sudden turn-up,
condition (\ref{eq:criterion}) breaks down, thus the
theory is inapplicable.
The measured cosmic errors, on the other hand,  are in accord with the
theory within the range of its validity. 
The agreement on small scales is 
even better for $\Delta S_k/S_k$
than for $\Delta F_k/F_k$, $k=3,4$. This, however, 
should not be over-interpreted,
as it is probably a coincidence due to cancellation effects
of the ratios $S_3={\bar \xi}_3/\xiav^2$ and
$S_4={\bar \xi}_4/\xiav^3$. 

The cosmic bias is always negative 
(right-hand panels of Fig.~\ref{fig:figure6}), 
i.e.~the biased estimators tend to
underestimate real values (SCB; HG). 
In this particular experiment, 
the measured cosmic bias is always dominated by
the measured cosmic error as predicted by the perturbative approach,
except for the largest scales. Here the cosmic bias
can become of same order as the cosmic error. 
HG suggested that the cosmic bias should be corrected for
when measuring cumulants. Whether this makes sense depends
on the magnitude of the {\em cosmic skewness}, 
i.e., the skewness of the cosmic
distribution function itself. This will be discussed in more
detail by paper II.  However, it is worth noting that 
function $\Upsilon(\tA)$ is positively skewed 
and that its maximum corresponds to the most likely measurement.
This  is in general 
smaller than the average, $\langle \tA \rangle$. 
Thus, as pointed out already by SC, the measured value $\tA$ 
in a finite sample is {\em likely} 
to underestimate the real value $A$ {\em even if $\tA$ 
is unbiased}. 
If the cosmic skewness and/or the cosmic variance are 
large compared to the
cosmic bias, it is pointless to correct for the cosmic bias. 
Either of the above is true for most surveys, 
including the upcoming wide-field surveys such as the 2dF and SDSS,
thus bias-corrected estimators are unlikely to be useful 
in the future.

%
%
\subsection{Cosmic Error and Cosmic Bias: Void Probability and 
Scaling Function}
\begin{figure}
\centerline{\hbox{\psfig{figure=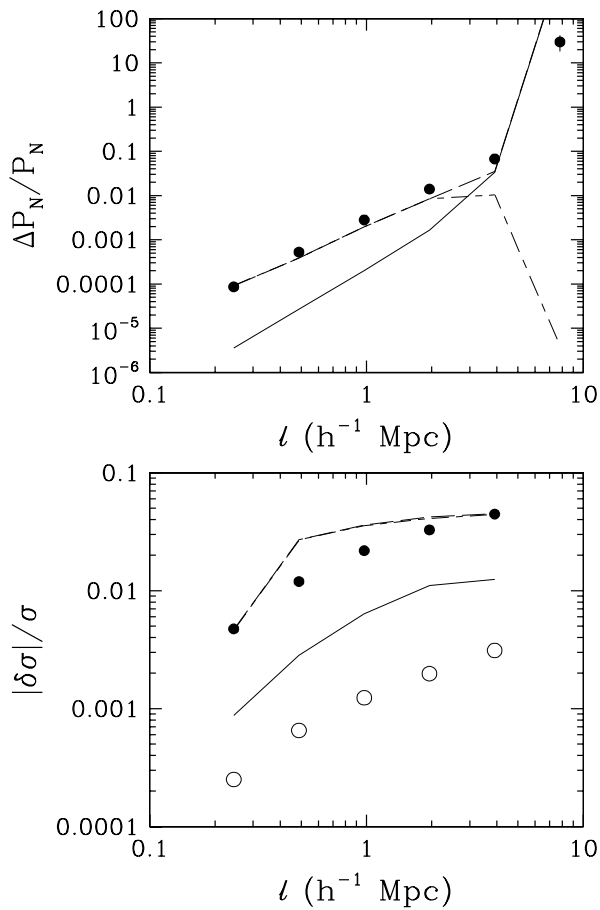,bbllx=167pt,bblly=285pt,bburx=450pt,bbury=708pt,width=8cm}}}
\caption[]{The cosmic error of the void probability function $P_0$ (upper panel) and
on the scaling function $\sigma=-\ln(P_0)/{\bar N}$ (lower panel). 
The measurements (filled symbols) are compared 
with the theoretical predictions of CBS
(long dashes). The finite volume error contribution is drawn 
with short dash-long dash, and the edge + discreteness effects 
contribution with solid lines. The
available scaling range is limited by the fact that on large scales
the measured void probability is zero. For $\ell \simeq 7.8
\ h^{-1}$ Mpc (upper right point on upper panel), the void probability
cancels from time to time in the subsamples ${\cal E}_i$. As a result,
it is possible to compute the unbiased function $\sigma$ but the 
estimated cosmic error on the biased estimator ${\tilde \sigma}$ is infinite. 
The open symbols in the lower panel  correspond to the measured
cosmic bias in $\sigma$.  It is positive and much smaller than
the cosmic error. It can be neglected for all 
the relevant dynamic range in the experiment considered here.}
\label{fig:figure7}
\end{figure}

%
%
Upper panel of Fig.~\ref{fig:figure7} shows $\Delta P_0/P_0$ as a function of
scale compared to the prediction of CBS (long dashes), with the finite
volume error contribution (dashes-long dashes) and with the edge + discreteness
contribution (solid curve). The agreement between theory and
prediction is excellent. 

The lower panel of Figure~\ref{fig:figure7} corresponds to
the scaling function $\sigma$. As for $\xiav$ and
$S_N$, the indicator ${\tilde \sigma}=-\ln(\tP_0)/\tilde {\bar N}$ 
is biased. This bias (open symbols) is of order 
$(\Delta\sigma/\sigma)^2$ and
can be neglected.\footnote{The theoretical and measured 
errors displayed on bottom part of Fig.~\ref{fig:figure7}
correspond to the biased indicator.}  The agreement between
theory and measurement is less impressive than for $P_0$, but 
this is mostly due to the difference of dynamic range covered
by the error in upper and lower panels of 
Fig.~\ref{fig:figure7}. 
Moreover, the calculation of $\Delta \sigma/\sigma$ by CBS 
is only approximate and could certainly be improved 
(see the discussion in CBS). 

The errorbars about $\sigma$ are quite small: nearly
an order of magnitude smaller than in Figs.~\ref{fig:figure4} and
\ref{fig:figure6}. 
According to equation~(\ref{eq:sigsig}), $\sigma$ reflects the 
low order statistics when $N_{\rm c}={\bar N}\, \xiav \ll 1$ ($\sigma
\la 1$ in Fig.~\ref{fig:figure3bis}) and the high 
order statistics when $N_{\rm c} \gg 1$ ($\sigma < 1$). From
the point of view of the errors, function $\sigma$ is
an excellent higher order indicator (as discussed earlier by CBS); it is
better than the low order factorial moments or cumulants, at least in
the non-linear regime $\ell \la \ell_0$. This fact alone unfortunately
does not guarantee the usefulness of this
statistic as various models of large scale structure formation 
could be degenerate with respect to the void probability.
The thorough work of Little \& Weinberg (1994) suggests 
that this is indeed the case.
It is tempting although dangerous to extrapolate the
results of their analysis to the function $\sigma$. 

%
%
\subsection{Cosmic Error: Counts-in-Cells}
\begin{figure}
\centerline{\hbox{\psfig{figure=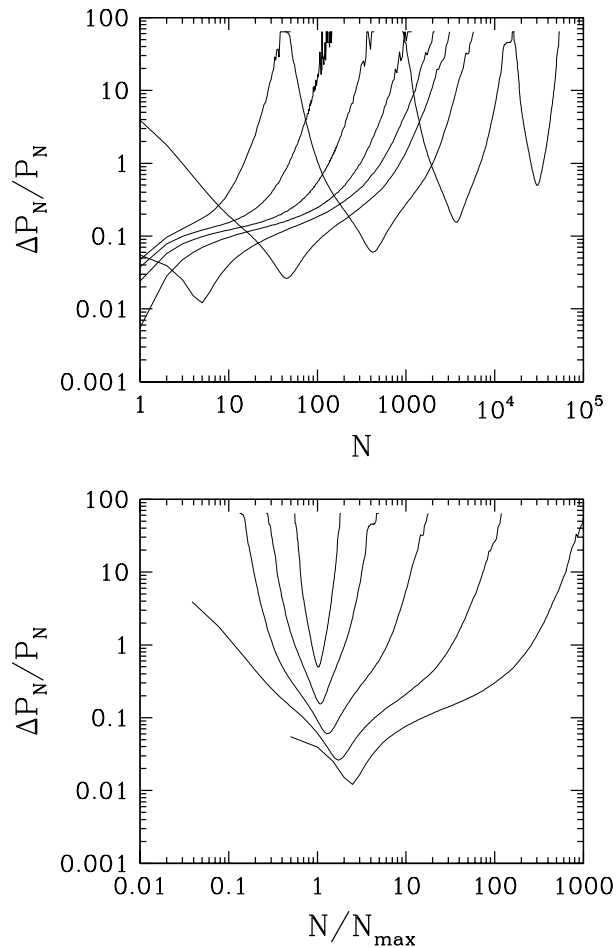,bbllx=172pt,bblly=267pt,bburx=450pt,bbury=704pt,width=8cm}}}
\caption[]{The cosmic error $\Delta P_N/P_N$ 
in the CPDF as a function of $N$ (upper panel) and as a function of
$N/N_{\rm max}$, where $N_{\rm max}$ is the value of $N$ for which
$P_N$ is maximum (lower panel). In the lower panel, only the scales
large enough so that $N_{\rm max} > 1$ are displayed.}
\label{fig:figure8}
\end{figure}

%
%
Upper panel of Fig.~\ref{fig:figure8} shows the cosmic error in the
CPDF as a function of $N$ for the various scales considered in ${\cal
E}_i$. The scale increases with the $x$-coordinate of the upper right
part of each curve. In the lower panel $\Delta P_N/P_N$ is
represented in a similar manner as a function of 
$N/N_{\rm max}$ where $N_{\rm max}$ is
the value of $N$ for which $P_N$ is a maximum. [We did not
display the (small) scales corresponding to $N_{\rm max}=0$ or $N_{\rm
max}=1$]. In agreement with intuition, 
the cosmic error reaches its minimum in the vicinity of $N
\simeq N_{\rm max}$ and becomes increasingly large in the tails. 
Thus the shape of the CPDF near its maximum has the most power to
constrain in terms of errors.
Kim \& Strauss (1998) have measured the cumulants 
$S_3$ and $S_4$ by fitting an
Edgeworth expansion convolved with a Poisson distribution 
to the measured CPDF in the 1.2 Jy IRAS galaxy catalog. 
According to their recipe, the best determined part
of the CPDF near the maximum was kept for the fit.
Their maximum likelihood approach
uses a simple model for the cosmic error, but their method is
promising. Its main weakness is the necessity to make a strong
prior assumption for the shape of the CPDF. A natural consequence is 
that the estimated errorbars on the measured cumulants are considerably
smaller than with the standard methods.

%
%
\subsection{Cosmic Correlations}
\begin{figure*}
\centerline{\hbox{
\psfig{figure=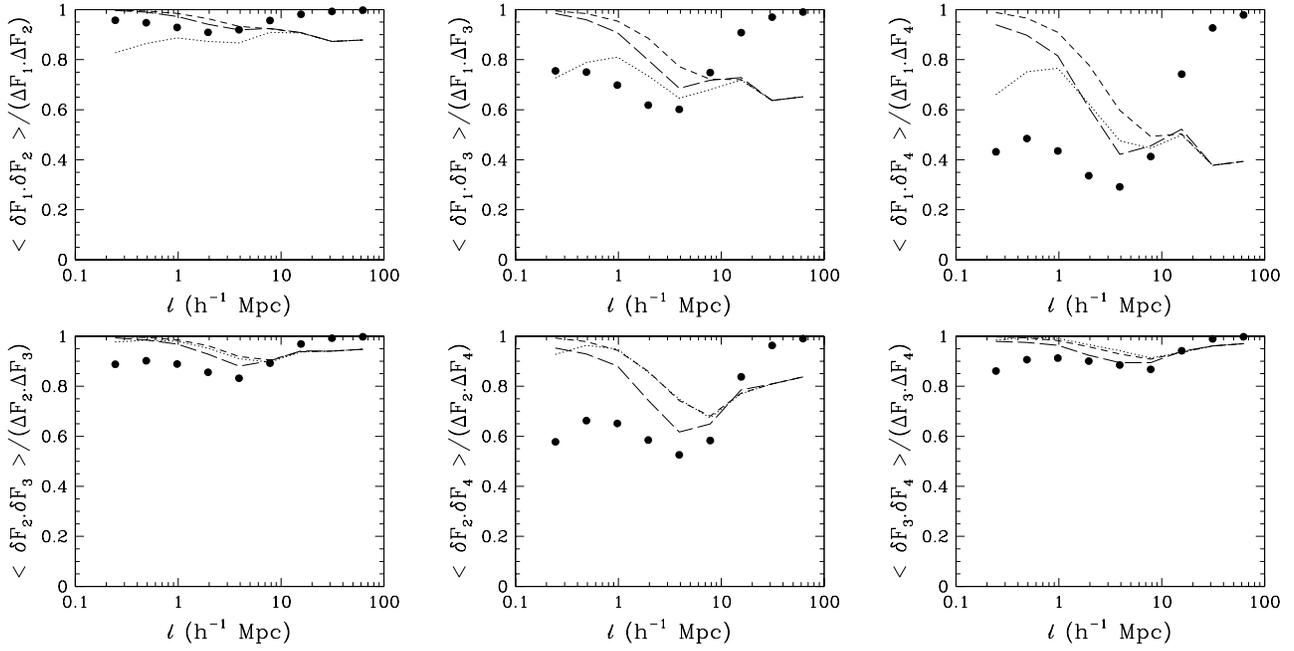,bbllx=69pt,bblly=447pt,bburx=576pt,bbury=708pt,width=17cm}}}
\caption[]{The measured cosmic cross correlation coefficients of the
factorial moments (symbols) are compared with
the models SS (dots), BeS (dashes), and E$^2$PT (long dashes).}
\label{fig:figure16}
\end{figure*}

\begin{figure*}
\centerline{\hbox{
\psfig{figure=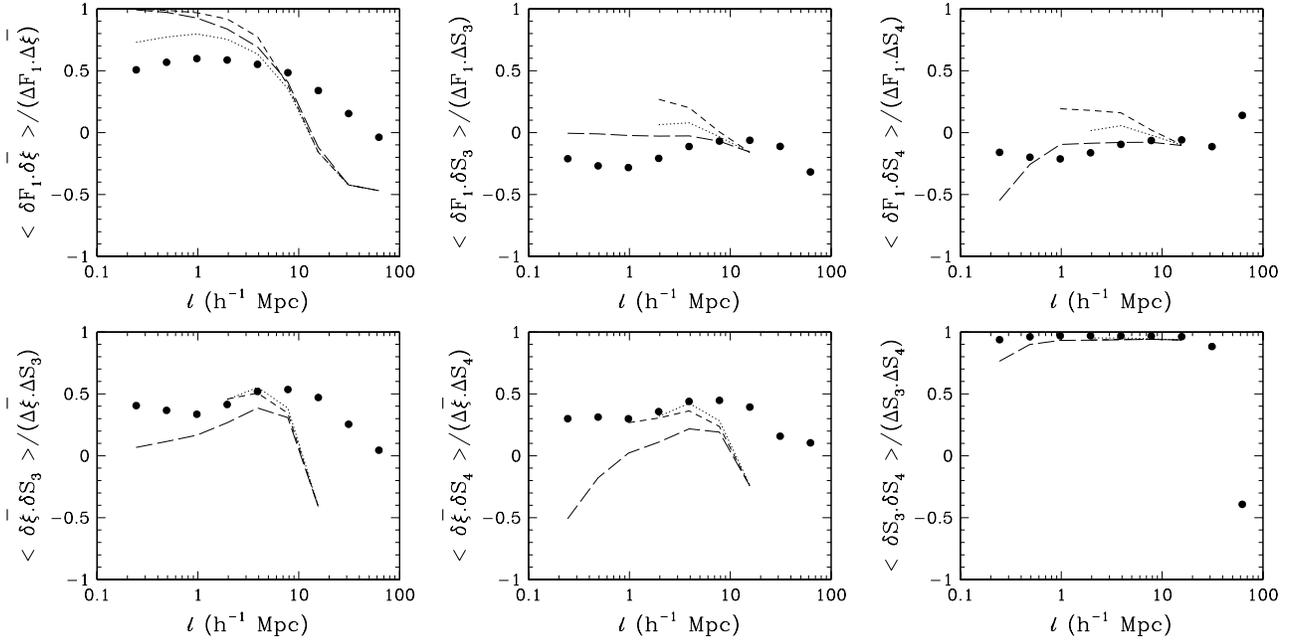,bbllx=88pt,bblly=446pt,bburx=540pt,bbury=672pt,width=17cm}
}}
\caption[]{Same as in Fig.~\ref{fig:figure16} but for
 the cosmic cross correlation coefficients of the
cumulants. The dynamic range for the theory is restrained by 
the condition~(\ref{eq:criterioneff}).}
\label{fig:figure17}
\end{figure*}

%
%
So far this section has dealt only with the second moment of the
cosmic distribution function, i.e. with the cosmic errors.
For a full description in the Gaussian limit, however, 
the moments of the joint distribution function are needed.
These moments form  the cosmic (cross-correlation) matrix (SCB).
It is defined as $\langle (\tA-\langle \tA \rangle)(\tB-\langle \tB \rangle) \rangle$, 
where $\tA$ and $\tB$ are any counts-in-cells related indicators,
for example $A=F_k(\ell)$ and $B=F_{k'}(\ell')$, 
or $A=\xiav(\ell)$ and $B=S_N(\ell')$, etc. 
A detailed theoretical investigation can be found in SCB (for $\ell=\ell'$). 
By definition, for two statistics $A$ and $B$, the
correlation coefficient $-1 \leq \rho \leq 1$ reads as
\begin{equation}
  \rho\equiv \frac{\langle \delta \tA \delta \tB \rangle}{\Delta A \Delta B}
      \equiv \frac{\langle (\tA-A)(\tB-B) \rangle}{\Delta A \Delta B}.
\label{eq:correcoef}
\end{equation}
The cosmic cross-correlation coefficient together with the errors
form the full correlation matrix. The inverse of this is the central
quantity for the joint probability distribution function in the
Gaussian limit.
As a preliminary numerical analysis, Figs.~\ref{fig:figure16} and
\ref{fig:figure17} present the correlation coefficients as functions of scale 
($\ell'=\ell$) for factorial moments and cumulants,  respectively.
As in Fig.~\ref{fig:figure4}, 
the dots, dashes and long dashes show the theoretical
predictions given by the SS, BeS and E$^2$PT models, 
respectively, as computed by SCB.
The computation of $\langle \delta \tA \delta \tB \rangle$ in
eq.~(\ref{eq:correcoef}) is analogous to that of
the cosmic error (see SCB for more details). 
[For $\Delta A$ and $\Delta B$, and
to have completely self-consistent calculations, 
we take the theoretical results as well
in eq.~(\ref{eq:correcoef})]. 

The agreement between theory and measurement is
less convincing for the cosmic cross-correlations 
than for the cosmic error. This appearance is 
due partly to the linear coordinates of the figures which
emphasize deviations, but nonetheless real.

On Fig.~\ref{fig:figure16} there is
a significant discord between theory and measurements for
factorial moments in the middle top, middle bottom, and top right panels.
On small scales, this result is quite natural: it is probably
due to the inaccuracy of the models SS, BeS and E$^2$PT 
employed to describe the underlying bivariate distributions 
(\S~4.2). In the weakly nonlinear regime, this
discrepancy is apparently puzzling, since the predicted 
cosmic error matches perfectly the
measurements (Fig.~\ref{fig:figure4}). The disagreement
increases with $|k-l|$, where $k$ and $l$ are the corresponding orders.
On large scales, the cross-correlations are dominated by
edge effects leading to the suspicion that the local Poisson
approximation (SC, \S~4.1) is becoming increasingly
inaccurate with $|k-l|$.\footnote{This is not surprising:
this approximation neglects local correlations. 
This is all the more inaccurate as the 
difference between the ``weights'' 
given to two overlapping cells, i.e. $(N)_k$ and
$(N)_l$ for factorial moments, increases.}  Another although
less likely possibility,
is that the leading order approach in $v/V$ is 
insufficient and higher order corrections are necessary to calculate
cross-correlations. It would go beyond the scope of this paper to analyse 
in detail these effects which are left for future research.

For the cumulants, in addition to the above arguments,
our perturbative approach to compute cross-correlation
allows only a narrow dynamic range for analytic predictions,
defined by criterion~(\ref{eq:criterion}). In Fig.~\ref{fig:figure17},
this condition is chosen for practical purpose to be
\begin{equation}
  |b_A| \leq \Delta A/A \leq 1.
  \label{eq:criterioneff}
\end{equation}
This is necessary but not sufficient:  the theory
appears to disagree significantly with 
the measurements on large scales at the top left, 
lower left and lower middle 
panels of Fig.~\ref{fig:figure17}. 

Despite some of the discrepancies, 
the general features of the cross-correlations are 
well described by the theoretical
predictions. For instance the cross-correlation between
two statistics $A_k$ and $A_l$ 
decreases  with the difference between the orders $|k-l|$ 
as predicted (SCB). In our 
particular experiment ${\bar N}$ is significantly correlated with
$\xiav$, but only weakly (anticorrelated) with
$S_k$, $k=3,4$. Similarly,
$\xiav$ and $S_3$ are weakly, but $S_3$ and $S_4$ are strongly correlated.
A detailed discussion on these effects can be found in SCB. 
%
%
%

%
%
\section{Summary and Discussion}
%
%
In this paper we have studied experimentally the properties of the moments
of the cosmic distribution  function of measurements $\Upsilon(\tA)$, 
where $\tA$ is an indicator of a  counts-in-cells statistic. 
For a thorough examination of $\Upsilon(\tA)$ itself the reader
is referred to paper II also in this volume.

We examined the factorial moments $F_k$, 
the cumulants $\xiav$ and $S_N$'s, the void probability $P_0$, 
its scaling function,  $\sigma\equiv -\ln(P_0)/F_1$, 
and the count-in-cells themselves $P_N$.
$\Upsilon(\tA)$ was measured in the largest available $\tau$CDM simulation
divided into 4096 cubical subsamples. In each of these many subsamples, 
$\tA$ was extracted, and its probability distribution 
function $\Upsilon$ was estimated with great accuracy. 
The main results of our analysis are the following:
\begin{enumerate}
\item The measured count-in-cells in the whole simulation, in particular 
the cumulants $S_N$, are in excellent agreement with perturbation theory 
predictions in the weakly nonlinear regime. This confirms the results of 
numerous previous investigations in an unprecedented dynamic range.
The textbook quality agreement demonstrates the 
state of the art accuracy of the simulation. Similarly,
the measurements confirm extended perturbation
theory (EPT) in the full available dynamic range $0.05 \la \xiav \la 50$, 
for $S_N$, $N \le 10$. In addition one loop perturbation theory predictions
based on the spherical model (Fosalba \& Gazta\~naga 1998) were found
to be an excellent description of the measured $S_N$ up to a 
${\bar \xi}\la 1$. 
\item The variance of $\Upsilon$ is the square of the 
expected cosmic error, $\Delta A$, 
in the measurement of $A$ in a subsample, identified with
a realization of the local observed universe. The measurement of 
$\Delta A/A$, for $A=P_0$, $\sigma$, $F_k$ and $S_N$ appears
to be globally in good accord with the 
theoretical predictions of Colombi, Bouchet \& Schaeffer (1995), 
Szapudi \& Colombi (1996, SC) and Szapudi, Bernardeau \& Colombi
(1998a, SCB).

In the  highly non-linear regime, the theoretical predictions of SC and
SCB tend to overestimate the cosmic error slightly, except for the ratios 
$S_3={\bar \xi}_3/\xiav^2$ and $S_4={\bar \xi}_4/\xiav^3$. In the latter case,
there are some cancellations and the agreement between theory 
and measurement is good even on small scales, but this is
probably a coincidence.
It appears thus that none of the three variants of the 
hierarchical model in SC and SCB, 
can give an accurate enough account of the non-linear behavior of gravitational 
dynamics for the bivariate distribution functions.\footnote{As discussed
in \S~4.1., the analysis
of the cosmic error indirectly probes the bivariate probability distribution 
function $P_{N,M}(r,\ell)$ of having $N$ and $M$ galaxies respectively 
in two cells of size $\ell$ separated by distance $r$ (see, e.g. SC).} 

In the weakly non-linear regime, agreement between theory and
predictions is excellent for the factorial moments, but less good
for the cumulants, due to the limitations of the perturbative
approach used to expand such ratios.

Nonetheless EPT yields the most precise overall agreement with theory
for our particular experiment.  On small scales $1\ h^{-1}$ Mpc 
$\la \ell \la 4\ h^{-1}$ Mpc, EPT overestimates the errors 
perhaps by a factor of two in the worst case.
\item In addition to the cosmic errors, the {\em
cosmic bias}, $b_A$, was studied in detail as well. An estimator is
biased when its ensemble average is different from the real
value: $b_A\equiv \langle\tA \rangle/A - 1 \neq 0$. This is always the
case when unbiased estimators are combined in a non-linear
fashion to form a new estimator (SCB, Hui \& Gazta\~naga 1998, HG), 
such as the cumulants.

In agreement with SCB, the measured cosmic bias
is of order $(\Delta A/A)^2$ and thus negligible when the cosmic error
is small. However, as for the errors, the theory tends to overestimate the
bias in the non-linear regime. On large scales, where the cosmic bias becomes
significant because of edge effects, 
the perturbative approach used by SBC to compute
theoretical predictions is then outside of its domain of validity.

Note that in the regime where the cosmic bias is significant, 
the cosmic error is likely to be large.
For instance, in the particular numerical experiment used in this paper, 
the cosmic bias was always smaller than the cosmic errors and in most
cases negligible. 
Moreover, in the regime where the bias could be significant,
the cosmic distribution function $\Upsilon(\tA)$ is significantly 
positively skewed (paper II). This implies that the
measured $\tA$ is likely to underestimate the true value even
for an unbiased estimator. The result is
an {\em effective} cosmic bias,  at most of order $\Delta A/A$. 
As already shown by SC, this effective bias can contaminate even
unbiased estimators such as
$\tF_k$ and $\tP_N$. As a consequence, it is pointless 
correcting for the cosmic bias,  in contrast with the proposition of HG,
unless it is done in the
framework of a maximum likelihood approach which takes
into account fully the effects of the shape of 
the cosmic distribution function. 

\item To complete the analysis of second moments, a preliminary
investigation of the cosmic correlation coefficients for
factorial moments and cumulants was conducted.
Together with the cosmic errors, these coefficients form the cosmic
cross-correlation matrix which underlies maximum likelihood
analysis in the Gaussian limit. 

Theoretical predictions of SBC give good qualitative account
of the measured correlation coefficients, although they become
increasingly approximate with the difference 
between the corresponding orders. This is likely to
be a consequence of the local Poisson assumption
(SC) employed for analytic predictions. 
\end{enumerate}
%

Provided that the Gaussian limit is reached in terms of the error
distribution, the formalism of SC and SBC allows for a maximum likelihood
analysis of the CPDF measured in
three-dimensional galaxy catalogs. Two preliminary investigations 
are currently being undertaken. 
Szapudi, Colombi \& Bernardeau (1999b) reanalyse already existing
joint measurements of $F_1$ and $\xiav$, and 
Bouchet, Colombi \& Szapudi (1999)
perform a likelihood analysis of the count-in-cells measured in the
1.2Jy IRAS survey (Bouchet et al.~1993). 

Paper II probes the domain of the Gaussian
approximation for the cosmic distribution function, together with 
preliminary investigations for the bivariate 
cosmic distributions $\Upsilon(\tA,\tB)$.
As shown there, the Gaussian limit is reached when the
relative cosmic error is small compared to unity. 
This is expected to hold for a large dynamic range in future large
galaxy surveys such as the 2dF and the SDSS (Colombi et al.~1998).

Statistical analyses of weak lensing surveys
are similar to counts-in-cells measurements
(e.g., Bernardeau, Van Waerbeke \& Mellier 1997; Mellier 1999; Jain,
Seljak \& White 1999). As a result, slight modification of the formalism
of SC and SCB is fruitful to compute theoretical cosmic errors
and cross-correlations (Bernardeau, Colombi \& Szapudi, 1999).

Finally, it is worth to mention a few questions
which were not addressed so far by the investigations presented
in this paper.
As light might not trace mass, the distribution of
galaxies may be biased (not to be confused with
the cosmic bias), and also realistic galaxy surveys are
subject to redshift distortion. While the above results were 
obtained for the mass, note that the theory which served
as a basis of comparison is quite general and was formulated
to describe phenomenologically either the mass or the
galaxies. It appears that there should be no 
qualitative changes introduced by biasing  
or redshift distortions (e.g., Szapudi et al., 1999e), 
thus the same theory can be used for the galaxies
as for the mass, except perhaps with slightly different
parameters, or underlying statistical models. In fact,
two of the models (SS, BeS) were entirely motivated by
the galaxy and not by the mass distribution; they are
expected to be more accurate for realistic
catalogs if used in a self-consistent fashion. The scaling
properties underlying these models is even more accurate
in redshift space, as is well known.
EPT, on the other hand, was originally motivated by theoretical
considerations of the mass distribution and numerical
simulations (Colombi et al.~1997), and therefore
it is no wonder that it is the most successful model for the
mass (but see also Scoccimarro \& Frieman 1998). 
Nonetheless, even EPT was found to be a fairly good
model for the galaxy distribution, at least in the EDSGC
survey (Szapudi, Meiksin \& Nichol 1996), a possible indication
that galaxies approximately trace mass after all. In addition, 
it is worth mentioning that biasing models can be nondeterministic,
i.e. stochastic in nature, but this again does not introduce
anything new qualitatively which could not be handled in the
framework of the theory of SCB. Finally, the theory outlined in 
this paper was contrasted against measurements in a $\tau$CDM 
simulation. However, the analytical framework is general enough to accommodate
any cosmological model, and there are no qualitative differences
in this respect between different cosmologies with Gaussian initial
conditions and hierarchical clustering. Thus
repeating the same analysis for a different CDM-like cosmogony
would be superfluous and inconsequential. 

\section*{Acknowledgments}
 The FORTRAN routine for computing $S_N$, $3 \leq N \leq 10$, using
 one loop perturbation theory predictions based on
 the spherical model was provided by P. Fosalba
 (see the right-hand panels of Fig.~\ref{fig:figure3}).
 We thank F. Bernardeau, P. Fosalba, C. Frenk, R. Scoccimarro,
 A. Szalay and S. White for useful discussions.
 It is a pleasure to acknowledge support for visits by IS and SC to the
 MPA, Garching and by SC to the dept Physics, Durham,
 during which part of this work was completed.           
 IS and AJ were supported by the PPARC rolling
 grant for
 Extragalactic Astronomy and Cosmology at Durham.

 The Hubble volume simulation data was made available by the Virgo
 Supercomputing Consortium
 (http://star-www.dur.ac.uk/$^{\sim}$frazerp/virgo/virgo.html).  The
 simulation
 was performed on the T3E at the Computing Centre of the Max-Planck
 Society in Garching. We would like to give our thanks to the many
 staff at the Rechenzentrum who have helped us to bring this project to
 fruition.  The FORCE package (FORtran for Cosmic Errors) used for the
 error-calculations in this paper is available
 on request from its authors SC and IS.

%

%
%

\appendix
\section{Definitions and Notations}
The count probability distribution function (CPDF) $P_N$, gives
the probability of finding $N$ objects in a cell of volume $v$
thrown at random in the catalog.

Factorial moments, $F_k$, are defined as follows
\begin{equation}
F_k \equiv \langle (N)_k \rangle \equiv \langle N(N-1)\cdots (N-k+1) \rangle=
\sum_N (N)_k P_N,
\label{eq:deffacmom}
\end{equation}
where the falling factorial $(N)_k$ is defined in the first part of the
equation.
The $F_k$ are proportional to the moments of the underlying 
density field
$\rho$ smoothed over the cell of volume $v$: $F_k = {\bar N}^k \langle \rho^k 
\rangle$ (SSa; assuming the normalization $\langle \rho \rangle=1$), 
where ${\bar N}$ is the average count in a cell:
\begin{equation}
{\bar N}\equiv \langle N \rangle = F_1.
\end{equation}

Counts-in-cells are related to quantities of 
dynamical interest,
such as the (connected) $N$-point correlation functions, $\xi_N$ (e.g., 
Peebles 1980). The averaged 
$N$-point correlation function over a cell is given by
\begin{equation}
{\bar \xi}_N \equiv \frac{1}{v^N} \int_v d^3 r_1 \cdots d^3 r_N \xi(r_1,\ldots,r_N).
\end{equation}
This is the connected moment of the smoothed
density field, ${\bar \xi}_N=\langle \delta^N \rangle_{\rm c}$ 
(with $\delta\equiv \rho-1$). The connected moments or cumulants
of a Gaussian field are identically zero for $N \geq 3$. 
In this paper, normalized cumulants are defined as
\begin{equation}
S_N\equiv \frac{{\bar \xi}_N}{{\bar \xi}^{N-1}},
\label{eq:cumu}
\end{equation}
with the short-hand notation ${\bar \xi}\equiv {\bar \xi}_2$.
By definition, $S_1\equiv S_2\equiv 1$, thus for second order
$\xiav$ is used.

The quantities $S_3$ and $S_4$ are often called skewness and kurtosis
in the astrophysical literature, although
their definition differs slightly from the original usage in statistics.
The reason for normalization in eq.~(\ref{eq:cumu}) is dynamical. The
$S_N$'s exhibit a weak scale dependence only due
to the scale-free nature of gravity.
In the highly nonlinear regime stable clustering is expected to set in, 
(e.g., Peebles 1980) and in the weakly nonlinear 
regime perturbation theory predicts approximate scaling depending
on the initial fluctuation spectrum
(e.g., Juszkiewicz, Bouchet \& Colombi 1993; Bernardeau 1994).

The counts-in-cells generating function, 
\begin{equation}
P(x) \equiv \sum_{N=0}^{\infty} x^N P_N,
\end{equation}
writes (White 1979; Balian \& Schaeffer 1989a; SSa)
\begin{equation}
P(x)=\exp\{-{\bar N}(1-x) \sigma[N_{\rm c} (1-x)] \},
\label{eq:pofx}
\end{equation}
where 
\begin{equation}
  N_{\rm c} \equiv {\bar N}\ \xiav
\end{equation}
is the typical number of objects in an overdense cell (e.g., Balian \&
Schaeffer 1989a), and
\begin{equation}
\sigma(N_{\rm c})=\sum_{N=1}^{\infty} (-1)^{N-1} \frac{S_N}{N!} N_{\rm c}^{N-1}.
\label{eq:sigofy}
\end{equation}
It is worth noticing that (White 1979; Balian \& Schaeffer 1989a; SSa)
\begin{equation}
P(x)=P_0[{\bar N}(1-x)],
\end{equation}
if the void probability is expressed in terms of average counts ${\bar N}$.
The measurement of $P_0$ is particularly interesting since it probes 
directly the count probability generating function:
\begin{equation}
  \sigma(N_{\rm c})=-\ln(P_0)/{\bar N}.
\end{equation}

The exponential generating function for factorial moments,
\begin{equation}
F(x)=\sum_{k\geq 0} F_k \frac{x^k}{k!}
\end{equation}
is directly related to $P(x)$ (SSa) through
\begin{equation}
F(x)=P(x+1).
\label{eq:fofx}
\end{equation}
Combining eqs.~(\ref{eq:pofx}), (\ref{eq:sigofy}) and (\ref{eq:fofx}),
one can obtain a useful relation between cumulants and factorial moments 
(SSa)
\begin{equation}
    S_N=\frac{\xiav F_N}{N_{\rm c}^N}-\frac{1}{N}\sum_{k=1}^{N-1}
    \left(\begin{array}{c} N \\ k \end{array} \right)
    \frac{(N-k)S_{N-k} F_k}{N_{\rm c}^k}.
   \label{eq:SqfromFk}
\end{equation}
The state of the art practical recipe consists of measuring the 
CDPF with high oversampling
(Sect.~3), computing the factorial moments from eq.~(\ref{eq:deffacmom}),
and finally calculating the cumulants from the above recursion
eq.~(\ref{eq:SqfromFk}). This procedure eliminates the need for
explicit discreteness correction.

\bsp
\label{lastpage}

%

\begin{thebibliography}{99}
%
\bibitem{} Balian R., Schaeffer R., 1989a, A\&A, 220, 1 
\bibitem{} Balian R., Schaeffer R., 1989b, A\&A, 226, 373 
\bibitem{b3} Baugh C.M., Gazta\~naga E.,  Efstathiou G., 1995, MNRAS, 274,
	1049
\bibitem{b5} Bernardeau F., 1994, A\&A, 291, 697
\bibitem{b6} Bernardeau F., 1996, A\&A, 312, 11  (B96)
\bibitem{} Bernardeau F., Colombi S., Szapudi I., 1999, in preparation
\bibitem{b7} Bernardeau F., Schaeffer R.,  1992, A\&A, 255, 1 (BeS)
\bibitem{} Bernardeau F., Van Waerbeke L., Mellier Y., 1997, A\&A, 322
\bibitem{b8} Bond J.R., Efstathiou G., 1984, ApJ, 285, L45
\bibitem{} Bouchet F.R., Colombi S., Szapudi I., 1999, in preparation
\bibitem{} Bouchet F.R., Schaeffer R.,  Davis M., 1991, ApJ, 383, 19 
\bibitem{b11} Bouchet F.R., Strauss M.A., Davis M., 
      Fisher K.B., Yahil A., Huchra J.P., 1993, ApJ, 417, 36
\bibitem{b14} Colombi S., Bernardeau F., Bouchet F.R., Hernquist L., 1997,
   MNRAS, 287, 241
\bibitem{b15} Colombi S., Bouchet F.R., Hernquist L., 1996, ApJ, 465, 14 (CBH)
\bibitem{b18} Colombi S., Bouchet F.R., Schaeffer R., 1992, A\&A, 263, 1 
\bibitem{b16} Colombi S., Bouchet F.R., Schaeffer R., 1994, A\&A, 281, 301
\bibitem{b17} Colombi S., Bouchet F.R., Schaeffer R., 1995, ApJS, 96, 401 (CBS)
\bibitem{b19} Colombi S., Charlot S., Devriendt J., Fioc M., Szapudi
I., 1999, in preparation 
\bibitem{} Colombi S., Szapudi I., Szalay A.S., 1998, MNRAS, 296, 253
\bibitem{} Efstathiou G., Davis M., Frenk C. S.,  White S. D. M., 1985,
ApJS, 57,241
\bibitem{} Eke V.R., Cole S., Frenk C.S., 1996, MNRAS, 282, 263
\bibitem{} Evrard A.E., et al., 1999, in preparation
\bibitem{} Fosalba P., Gazta\~naga E., 1998, MNRAS, 301, 503
\bibitem{b32} Gazta\~naga E., Baugh C.M., 1995, MNRAS, 273, L1 
\bibitem{b1}  Hamilton A.J.S., Kumar P., Lu E., Matthews A., 1991,
ApJ, 374, L1
\bibitem{} Hockney R. W., Eastwood J. W., 1981, Computer simulation using particles
(McGraw Hill)
\bibitem{}  Hoyle, F., Szapudi, I., Baugh, C.M., MNRAS, in prep.
\bibitem{b3}  Hui L., Gazta\~naga E., 1998, preprint (astro-ph/9810194) (HG)
\bibitem{} Jain B., Seljak U., White S.D.M., 1999, preprint (astro-ph/9901191)
\bibitem{} Jenkins A.,  et al., 1998, ApJ, 499, 20
\bibitem{b40} Juszkiewicz R., Bouchet F.R., Colombi S., 1993, ApJ, 412, L9
\bibitem{b41} Juszkiewicz R., Weinberg D.H., Amsterdamski P., Chodorowski M.,
   Bouchet F.R., 1995, ApJ, 442, 39
\bibitem{} Kerscher, M., Szapudi, I. \& Szalay, A.S., 1999, ApJ, in prep.
\bibitem{b45} Kim R.S., Strauss M.A., 1998, ApJ, 493, 39
\bibitem{} Little B., Weinberg D.H.,1994, MNRAS, 267, 605
\bibitem{} Lokas E.L., Juszkiewicz R., Bouchet F.R., Hivon E., 1996, ApJ, 467, 1
\bibitem{} MacFarland T., Couchman H. M. P., Pearce F. R., Pichlmeier J.,
1998, New Astronomy, Vol 3, no.8, 687

\bibitem{} Mellier Y., 1999, Ann. Rev. Astron. Astrophys., in press 
\bibitem{b2}  Peacock J.A., Dodds S.J., 1996, MNRAS, 280, 19P (PD)
\bibitem{} Peebles, P.J.E., 1980,  The Large-Scale
Structure of the Universe (Princeton University Press, 1980), p.~147, 149, 150  
\bibitem{b4}  Scoccimarro R., 1998, MNRAS, 299, 1097
\bibitem{b64} Scoccimarro R., Colombi S., Fry J.N., Frieman J.A., Hivon E.,
   Melott A., 1998, ApJ, 496, 586
\bibitem{} Scoccimarro R., Frieman J.A., 1998, preprint 
   (CITA-98-17, FERMILAB-Pub-98/255-A)
\bibitem{b66} Szapudi I., 1998, ApJ, 497, 16
\bibitem{b69} Szapudi I., Colombi S., Bernardeau F., 1999a, submitted
to MNRAS (SCB)
\bibitem{} Szapudi I., Colombi S., Bernardeau F., 1999b, in
preparation
\bibitem{} Szapudi I., Colombi S., Jenkins A., Colberg J., 1999c, submitted to MNRAS (paper II)
\bibitem{} Szapudi I., Quinn T., Stadel J., Lake G., 1999d, Apj, 517, 54
\bibitem{} Szapudi I., et al., 1999e, in preparation
\bibitem{b68} Szapudi I., Colombi S., 1996, ApJ, 470, 131 (SC)
\bibitem{b70} Szapudi I., Meiksin A., Nichol R., 1996, ApJ, 473, 15
\bibitem{ssb92}  Szapudi, I., Szalay, A.S., \& Bosch\'an, P.,
1992, Apj, 390, 350               
\bibitem{b71} Szapudi I., Szalay A.S., 1993a, ApJ, 408, 43  (SSa)
\bibitem{b72} Szapudi I., Szalay A.S., 1993b, ApJ, 414, 493 
\bibitem{b74} Szapudi I., Szalay A.S., 1997, ApJ, 481, L1
\bibitem{} White S.D.M., 1979, MNRAS, 186, 145 
\bibitem{} White S.D.M., 1996, in Cosmology and Large-Scale Structure,
eds. R. Schaeffer, J. Silk, M. Spiro \& J. Zinn-Justin (Dordrecht: Elsevier)
\bibitem{} White S.D.M., Efstathiou G., Frenk C.S., 1993, MNRAS, 262, 1023
\bibitem{} Zel'dovich Ya.B., 1970, A\&A, 5, 84

\end{thebibliography}
\end{document}